\documentclass[prd,jmp,reprint,onecolumn,superscriptaddress,12pt]{revtex4-2}

\usepackage{bm}
\usepackage{amsfonts}
\usepackage{latexsym}
\usepackage[latin1]{inputenc}
\usepackage{graphicx}
\usepackage{amsmath}
\usepackage[mathscr]{eucal}
\usepackage{palatino}
\usepackage{mathpazo}
\usepackage{textcomp}
\usepackage{float}
\usepackage{booktabs}
\usepackage{dcolumn}
\usepackage{ragged2e}
\usepackage{hyperref}
\hypersetup{colorlinks,citecolor=blue}
\hypersetup{colorlinks=true,linkcolor=red,filecolor=magenta,    urlcolor=blue}
\usepackage{amsmath}
\usepackage{xcolor}
\usepackage{orcidlink}
\usepackage[caption=false]{subfig}
\usepackage{commath}
\begin{document}

\title{Magnetically charged regular black holes in $f(R,T)$ gravity coupled to nonlinear electrodynamics}

\author{Takol Tangphati}
\email{takoltang@gmail.com}
\affiliation{School of Science, Walailak University, Thasala, Nakhon Si Thammarat, 80160, Thailand.}

\author{Menglong Youk}
\email{menglongyouk@yahoo.com}
\affiliation{Strong Gravity Group, Department of Physics, Faculty of Science, Silpakorn University, Nakhon Pathom 73000, Thailand}

\author{Supakchai Ponglertsakul}
\email{supakchai.p@gmail.com}
\affiliation{Strong Gravity Group, Department of Physics, Faculty of Science, Silpakorn University, Nakhon Pathom 73000, Thailand}

\date{\today}

\begin{abstract}
We construct asymptotically flat, static spherically symmetric black holes with regular centre in $f(R,T)$ gravity coupled to nonlinear electrodynamics Lagrangian. We obtain generalized metric function of the Bardeen and Hayward black holes. The null, weak and strong energy conditions of these solutions are discussed. All the energy conditions hold outside the black hole's outer event horizon by appropriated choices of parameters. Quasinormal mode of massive scalar perturbation is also investigated. Quasinormal frequencies are computed via the sixth order Wentzel-Kramers-Brillouin (WKB) with Pad\'e approximation. All the imaginary parts of the frequencies are found to be negative. Finally, we provide an analysis in the eikonal limit.
\end{abstract}

\maketitle
\section{Introduction}

The most well-known  gravitational theory describing the relation between spacetime and matter is Einstein's general relativity (GR). For over a decade, this theory has been well-tested by the observations and experiments in the weak field limit like our solar system, and the highly dense binary systems \cite{Nojiri:2010wj,Coley:2019tyx}. However, there are numerous open questions that GR fails to provide answers, for instance, an accelerated expansion of the universe \cite{SupernovaSearchTeam:1998fmf,SupernovaCosmologyProject:1998vns}, and galaxy rotation curve \cite{Rubin:1970zza}. Rather than applying auxiliary fields to the theory, one could construct the modification of the GR as the extension based on the original Einstein's general relativity. 

One of the modifications of GR is the $f(R)$ gravity where the Ricci scalar $R$ in the Einstein-Hilbert action is replaced with an arbitrary function of $R$ \cite{RevModPhys.82.451,DeFelice:2010aj}. This modification can describe the accelerated expansion of the Universe without relying on the exotic matters \cite{Nojiri:2017ncd}. Moreover, the generalization of the $f(R)$ gravity theory leads to extra degrees of freedom related to curvature invariants and scalar fields, which are called Extended Theories of Gravity (ETG) \cite{Capozziello:2014bqa}. These additional degrees of freedom play a major role as effective fluids unlike the fluids of ordinary matter which is adopted as sources of the field equations. One class of the ETG is the $f(T)$ gravity theory where the extension of the torsional gravity with an arbitrary function of the torsion scalar $f(T)$ plays a major role in explaining the cosmological and astrophysics problems \cite{Cai:2015emx}. Additionally, another class of the extension of the Einstein's gravity theory is the $f(Q)$ gravity theory constructed from the symmetric teleparallel gravity which is based on the non-metricity scalar $Q$. The modification of this theory represents the stable dark energy causing the accelerated universe in which the matter perturbation remains constant \cite{BeltranJimenez:2017tkd, Khyllep:2022spx}.

In addition, the $f(R,T)$ gravity theory is designed to add the matter components into the gravitational action by applying the arbitrary function of the Ricci scalar $R$ along with the trace of the energy momentum tensor $T$. This is proposed in \cite{Harko:2011kv} where the modified field equation is derived and cosmological solution is analysed by introducing a self-interacting scalar field.  Numerous works of the $f(R,T)$ gravity theory have been investigated. The cosmological solutions based on a homogeneous and isotropic spacetime through a phase-space analysis are done in \cite{Shabani:2013djy}. In addition, several cosmological solutions from the $f(R,T)$ theory have been exclusively explored in refs \cite{Houndjo:2011fb,Jamil_2012,Baffou:2013dpa,Sharif:2014ioa,Mishra_2018}. The violation of the energy conditions is investigated in \cite{Alvarenga:2012bt,Sharif:2012ce}. Moreover, thermodynamics properties of the the $f(R,T)$ gravity theory are explored in \cite{Sharif:2012zzd,Jamil:2012ti,Jamil:2012pf,Sharif:2013ffa}. On the other hand, within the $f(R,T)$ framework, various compact objects are constructed and studied e.g., wormhole \cite{Banerjee:2023vdd} and compact stars \cite{Tangphati:2022mur,Tangphati:2023efy}.

Black holes are ones of the most fundamental objects in the universe. They play a crucial role in almost all relativistic gravitational field theories. The detection of gravitational waves \cite{LIGOScientific:2016aoc} and the first capture of black hole's image \cite{Akiyama_2019} marked the beginning of black hole's astronomy era. This makes black holes extremely important in astrophysics research nowadays. Black holes are solutions of relativistic gravitational field equations. According to GR, there is an essential singularity hidden behind each black hole. The regular black hole proposed by Bardeen \cite{bardeen1968non} offers a new possibility to obtain black hole without a singularity. Later, it is shown that regular black holes are the solutions of Einstein's gravity coupled to nonlinear electrodynamics \cite{Ayon-Beato:1998hmi} and the Bardeen black hole can be regarded as a nonlinear magnetic monopole \cite{Ayon-Beato:2000mjt}. The Bardeen black hole is later extended to include a cosmological constant \cite{Fernando:2016ksb}. Charged regular black holes with various mass functions are studied in \cite{Balart:2014cga}. In addition, a modification of the Reissner-Nordstr\"om black hole yields regular charged black hole, and its entropy obeys Bekenstein's area law \cite{Culetu:2014lca}. We refer interested readers to ref \cite{Bronnikov:2022ofk} for a recent review on regular black holes with nonlinear electrodynamics sources. Beyond GR, the regular black holes with nonlinear electrodynamics are extensively explored e.g. in Einstein-Gauss-Bonnet theory \cite{PhysRevD.97.104050,Chatterjee:2021ops,Kumar:2020uyz} and $f(R)$ gravity \cite{Rodrigues:2015ayd}.

In $f(R,T)$ gravity, an exact black hole solution surrounded by anisotropic fluid is explored \cite{Santos:2023fgd}. The energy conditions for each particular equation of state parameter $w$ are discussed in \cite{Santos:2023fgd}. This prompts a research question whether there are other black hole solutions in the $f(R,T)$ gravity. Thus, in this work, we construct asymptotically flat, static spherically symmetric regular black holes within the framework of $f(R,T)$ gravity. There are two approaches to obtain the black hole solutions. Firstly, we specifically choose a mass function that yields a regular black hole, and find the corresponding the nonlinear electrodynamics Lagrangian ($L_{NED}$). Secondly, we specify the $L_{NED}$, and find the corresponding mass function. From both approaches, we obtain novel magnetically charged regular black holes. Remarkably, from the second approach, we obtain a metric function that can be considered as a generalization of Bardeen and Hayward black hole \cite{Ayon-Beato:2000mjt,Hayward:2005gi,Fan:2016hvf}. Then, we analyse the null, weak and strong energy conditions of these solutions. The quasinormal modes and the eikonal limit of these black holes are also investigated.

This paper is organized as follows. In Sec~\ref{sec:Basiceq}, we discuss the $f(R,T)$ gravity coupled to nonlinear electrodynamics. The modified field equation is derived and corresponding energy-momentum tensor is given. Then, modified field equations are solved and the regular black holes are explored in Sec~\ref{sec:modified}. Then, we discuss the energy conditions in Sec~\ref{sec:energycond}
. We study quasinormal modes and the eikonal limit in Sec~\ref{sec:QNMs}. Lastly, we summarize our results and discuss possible extensions of this work in Sec~\ref{sec:conclud}.


\section{Basic equations}\label{sec:Basiceq}
We consider $f(R,T)$ gravity coupled to nonlinear electrodynamics (NED). This theory is described by 
\begin{align}
    S &= \frac{1}{2}\int\sqrt{-g}d^4 x~f(R,T) + \int\sqrt{-g} d^4 x~L_{NED},
\end{align}
where $f(R,T)$ is an arbitrary function of the Ricci scalar $R$ and the trace $T$ of the energy-momentum tensor of the matter $T_{\mu\nu}$. The nonlinear electrodynamics Lagrangian is given by $L_{NED}(F)$ where $F=-\frac{1}{4}F_{\mu\nu}F^{\mu\nu}$. The Faraday-Maxwell tensor is defined in term of the gauge potential $F_{\mu\nu}=\partial_{\mu}A_{\nu} - \partial_{\nu}A_{\mu}$.

Varying this action with respect to $\delta g^{\mu\nu}$, yields the modified Einstein field equation
\begin{align}
   \mathcal{G}_{\mu\nu} \equiv f_R R_{\mu\nu} + \left(g_{\mu\nu}\Box-\nabla_{\mu}\nabla_{\nu}\right)f_R - \frac{1}{2}f g_{\mu\nu} &=  T_{\mu\nu} - f_T (T_{\mu\nu} + \Theta_{\mu\nu}), \label{EFE}
\end{align}
where $f_R = \frac{\partial f}{\partial R}, f_T = \frac{\partial f}{\partial T}$ and $\Box = \nabla_{\alpha}\nabla^{\alpha}$. The energy-momentum tensor $T_{\mu\nu}$ and $\theta_{\mu\nu}$ are computed from
\begin{align}
    T_{\mu\nu} &\equiv -\frac{2}{\sqrt{-g}}\frac{\delta\left(\sqrt{-g}L_{NED}\right)}{\delta g^{\mu\nu}}, \\
    \theta_{\mu\nu} &\equiv g^{\alpha\beta}\frac{\delta T_{\alpha\beta}}{\delta g^{\mu\nu}}.
\end{align}
With nonlinear electrodynamics sources, the explicit forms of $T_{\mu\nu}$ and $\Theta_{\mu\nu}$ are expressed as
\begin{align}
    T_{\mu\nu} &= g_{\mu\nu}L_{NED} + L_F F_{\mu\gamma}{F_{\nu}}^{\gamma}, \\
    \Theta_{\mu\nu} &= -g_{\mu\nu}L_{NED} - F_{\mu\gamma}{F_{\nu}}^{\gamma}\left[\frac{L_{FF}}{2}F_{\rho\sigma}F^{\rho\sigma}+L_F\right],
\end{align}
where $L_F = \partial L_{NED}/\partial F$ and $L_{FF} = \partial^2 L_{NED}/\partial F^2$. Moreover, taking the trace of $\eqref{EFE}$ gives the following 
\begin{align}
    \Box f_R &= \frac{1}{3}\left( T - f_T (T+\Theta) + 2 f - f_R R\right), \label{traceEFE}
\end{align}
where $T\equiv g_{\mu\nu}T^{\mu\nu}$ and $\Theta \equiv g_{\mu\nu}\Theta^{\mu\nu}$. 
The equation of motion of the gauge field is 
\begin{align}
     \partial_{\mu}\left[\sqrt{-g}\left(4f_T L_{FF}F-L_F \right) F^{\mu\nu}\right] &=0. \label{NEDeq}
\end{align}
Now, we consider a static spherically symmetric solution. The line element written in Schwarzschild-like coordinate is given by
\begin{align}
    ds^2 = - A(r) dt^2 + B(r) dr^2 + r^2\left(d\theta^2 + \sin^2\theta d\phi^2\right). \label{lineelement}
\end{align}
We also consider a purely magnetic ansatz of the Faraday-Maxwell tensor \cite{Ayon-Beato:2000mjt}
\begin{align}
    F^{\theta\phi} &= \frac{q_m \csc \theta}{r^4}, \label{Fanswer1}
\end{align}
where $q_m$ is an integration constant that can be interpreted as the magnetic charge of the source. With this choice, the invariant $F$ is $-\frac{q_m^2}{2r^4}$. One can show that the ansatz \eqref{Fanswer1} satisfies the equation of motion \eqref{NEDeq}.


\section{Solving the modified field equations}\label{sec:modified}
Here we consider the $f(R,T) = R + 2 \beta T$ where $\beta$ is an  arbitrary constant. Together with purely magnetic field strength \eqref{Fanswer1}, the modified field equations i.e., $\mathcal{G}{_\mu}^{\nu} = T{_\mu}^{\nu} - f_T \left( T{_\mu}^{\nu} + \Theta{_\mu}^{\nu}\right)$, are 
\begin{align}
    -\frac{1}{r^2} + \frac{1}{B r^2} - \frac{B'}{B^2r} &=\left(1+4\beta\right) L_{NED} + \frac{2\beta q_m^2}{r^4}L_{F}, \\
    -\frac{1}{r^2} + \frac{1}{B r^2} - \frac{A'}{AB^2r} &=\left(1+4\beta\right) L_{NED} + \frac{2\beta q_m^2}{r^4}L_{F}, \\
    \frac{A''}{2AB} - \frac{A'B'}{4AB^2} + \frac{A'}{2ABr} - \frac{A'^2}{4A^2B} - \frac{B'}{2B^2r} &= \left(1+4\beta\right) L_{NED} + \frac{q_m^2} {r^4}\left(1+2\beta\right)L_F + \frac{2\beta q_m^4}{r^8}L_{FF},
\end{align}
where prime denotes derivative with respect to $r$. The first two equations imply that $A=B^{-1}$. Therefore, the remaining field equations are 
\begin{align}
    \frac{A'}{r} + \frac{A}{r^2} - \frac{1}{r^2} &= \left(1 + 4\beta\right)L_{NED} + \frac{2\beta q_m^2}{r^4}L_F, \label{EFE00}\\
A'' + \frac{2A'}{r} &=  2\left(1+4\beta\right)L_{NED} + \frac{2q_m^2}{r^4}\left(1+2\beta\right)L_F + \frac{4\beta q_m^4}{r^8}L_{FF} \label{EFE22},
\end{align}
In addition, the Ricci scalar is 
\begin{align}
    R &= -\left(A'' + \frac{4A'}{r} + \frac{2A}{r^2} - \frac{2}{r^2}\right).
\end{align}
Substituting this into the trace of the modified field equations \eqref{traceEFE} allows us to eliminate $A''$ in \eqref{EFE22}. After eliminating $A''$, we find that \eqref{EFE00} and \eqref{EFE22} are identical. Thus, we are left with a single first order ordinary differential equation (recalls that $F=F(r)$)

\begin{align}
   m'(r) &= -\frac{r^2}{2}\left(1+4\beta\right)L_{NED} - \frac{\beta q_m^2}{r^2}L_F,
   \label{Field_equation_1}
\end{align}
where the mass function $m(r)$ is defined via $A(r) \equiv 1- 2m(r)/r$. There are two ways to solve this equation. Firstly, we may choose a particular form of $m(r)$, then solving \eqref{Field_equation_1} for $L_{NED}$. Secondly, we fix the form of NED Lagrangian and solve for the mass function $m(r)$.

Before solving for a new solution, let's examine the consistency of \eqref{Field_equation_1}. We consider the case where $L_{NED}=-F, L_F = -1$. Therefore \eqref{Field_equation_1} can be solved as
\begin{align}
    m(r) &= -2M + \frac{q_m^2}{4r},
\end{align}
where $M$ is an integration constant. By letting $q_m = 2Q_m$, we obtain a special case of dyonic Reissner-Nordstr\"om black hole \cite{PhysRevD.87.081505}. We remark that when the Lagrangian matter reduces to the $U(1)$ electromagnetic, $\beta$ automatically disappears from $f(R,T) = R + 2\beta T$ since the energy-momentum tensor is traceless.



\subsection{Fixed mass function}

In this subsection, we shall solve the modified field equation \eqref{Field_equation_1} for the spherically symmetric regular black hole solution. We choose the mass function to be in the form
\begin{align}
    m(r) = M e^{-\frac{q_m^2}{2Mr}}, \label{massI}.
\end{align}
Here $M$ is a constant parameter and $q_m$ is the charge of the regular black hole. We remark that this form of mass function is considered to obtain regular black holes within the context of GR \cite{Culetu:2013fsa, Balart:2014cga, Culetu:2014lca} and $f(R)$ \cite{Rodrigues:2015ayd} gravity coupled to NED.

\begin{figure}[ht]
    \centering
    \includegraphics[width = 9cm]{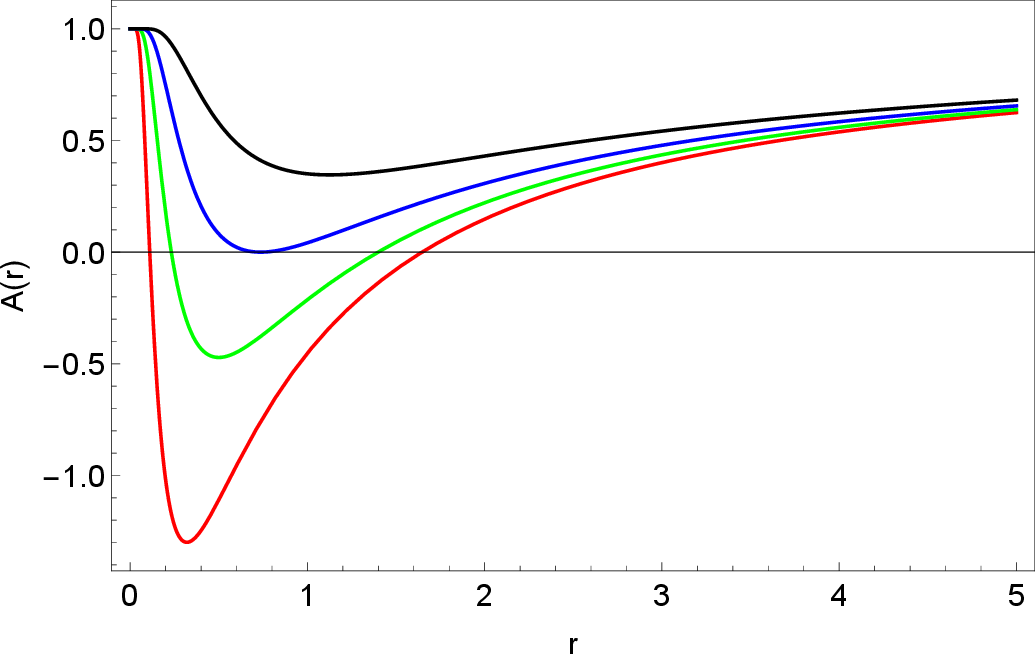}
    \caption{The metric function $A$ is plotted against radial coordinate $r$. The location of zeros indicates the location of black hole's event horizons for $M=1$ and $q_m=0.8$ (red), $q_m=1$ (green), $q_m=1.213$ (blue) and $q_m=1.5$ (black).}
    \label{fig:plotA}
\end{figure}

\begin{figure}[ht]
    \centering
    \includegraphics[width = 8cm]{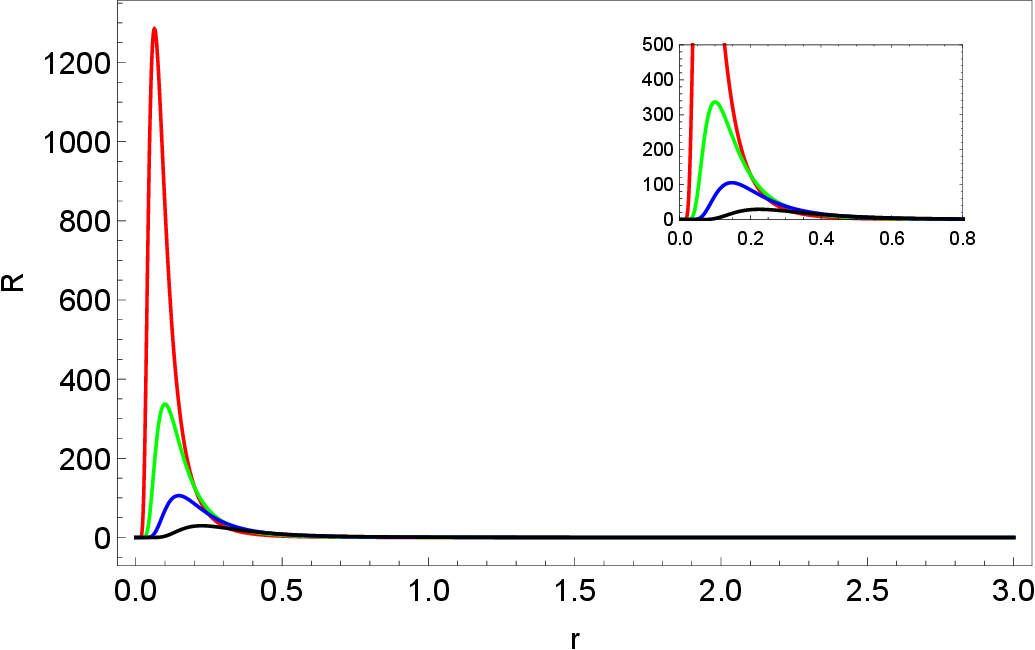}
    \includegraphics[width = 8.2cm]{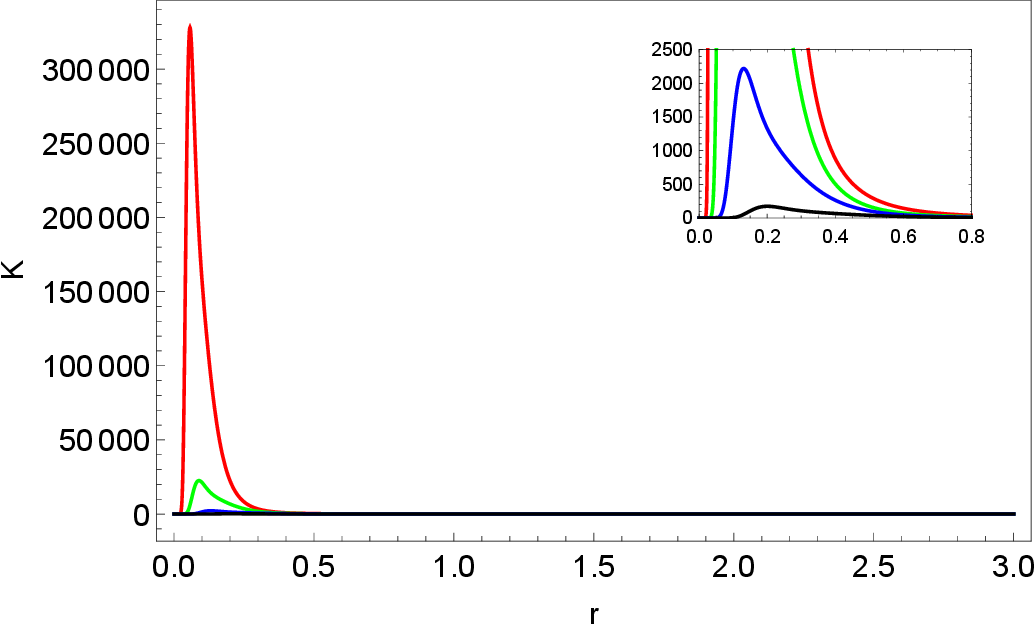}
    \caption{Left: The Ricci scalar, Right: The Kretechmann scalar for $M=1$, $q_m=0.8$ (red), $q_m=1$ (green), $q_m=1.213$ (blue) and $q_m=1.5$ (Black). The sub-figures show the behaviour of $R$ and $K$ at small $r$.  }
    \label{fig:scalarcurvature}
\end{figure}

Since~ $\displaystyle {\lim_{r\to\infty} \frac{m(r)}{r} = M} $, one can interpret the constant $M$ as the black hole's mass. The black hole's event horizons are determined by $A(r_h) =0$, and the location of the outer event horizon is given by
\begin{align}
    r_h &= - \frac{q_m^2}{2M \Omega(-\frac{q_m^2}{4M^2})}, \label{horizonmassI}
\end{align}
where $\Omega(z)$ is the omega function or the Lambert W function. This mass function \eqref{massI} allows for three possible outcomes regarding number of the horizons, i.e., two positive real roots (inner and outer horizon), one degenerated root (extremal case) and no real root (horizonless case). Remark that, throughout this work, we shall particularly focus on the first two cases. The behaviour of $A(r)$ are shown in Fig~\ref{fig:plotA} for three possible solutions. It can be observed from the figure that at small $r$, these solutions are finite i.e., $A(r) \sim 1$. The solutions are clearly asymptotically flat since $A(r) \to 1$ as $r\to \infty$. As $q_m$ increases, the minimum value of $A$ increases until $A(r)>0$ for all $r$.
The regularity of the solution can be observed by considering two curvature scalar quantities, i.e., the Ricci scalar $R$ and the Kretchmann scalar $K$. For the mass function \eqref{massI}, we obtain
\begin{align}
    R &= \frac{e^{-\frac{q_m^2}{2Mr}}q_m^4}{2Mr^5}, \\
    K &= R_{\mu\nu\sigma\rho}R^{\mu\nu\sigma\rho} = \frac{e^{-\frac{q_m^2}{Mr}}}{4M^2r^{10}}\left(q_m^8 - 16Mq_m^6r + 96M^2q_m^4r^2 - 192M^3q_m^2r^3 + 192M^4r^4\right).
\end{align}
In Fig~\ref{fig:scalarcurvature}, we display example plots of the Ricci and the Kretchmann scalars. The curvature scalars are finite everywhere for various values of $q_m$. Moreover, $R$ and $K$ behave as  $O(r^{-5}),O(r^{-6})$, respectively as $r\to\infty.$ In addition, the maximum value of $R$ locates at $r=\frac{q_m^2}{10M}$. On the other hand, the radius renders $K_{max}$ is not trivial. For instance, when $M=1$ and $q_m=0.8$, $R_{max}$ is $1,285$ at $r=0.064$ while $K_{max}$ is $327,982$ at $r=0.057.$ We emphasize that the mass function \eqref{massI}, together with these scalar curvatures, is already considered in GR and $f(R)$ gravity coupled to NED 
\cite{Balart:2014cga,Rodrigues:2015ayd}.

\begin{figure}[ht]
    \centering
    \includegraphics[width = 8.35cm]{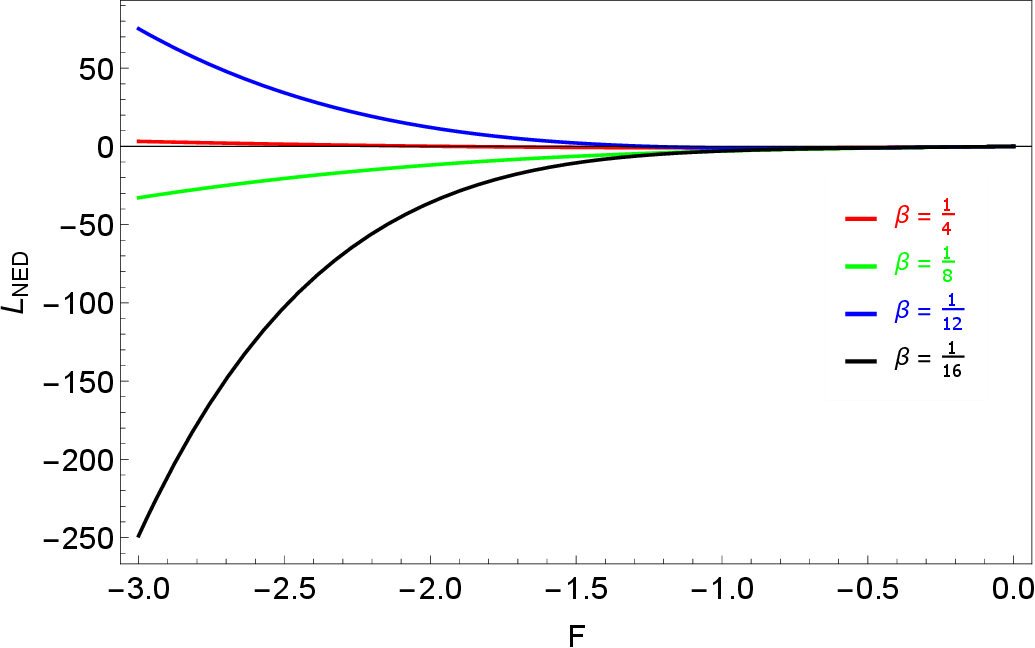}
    \includegraphics[width = 8cm]{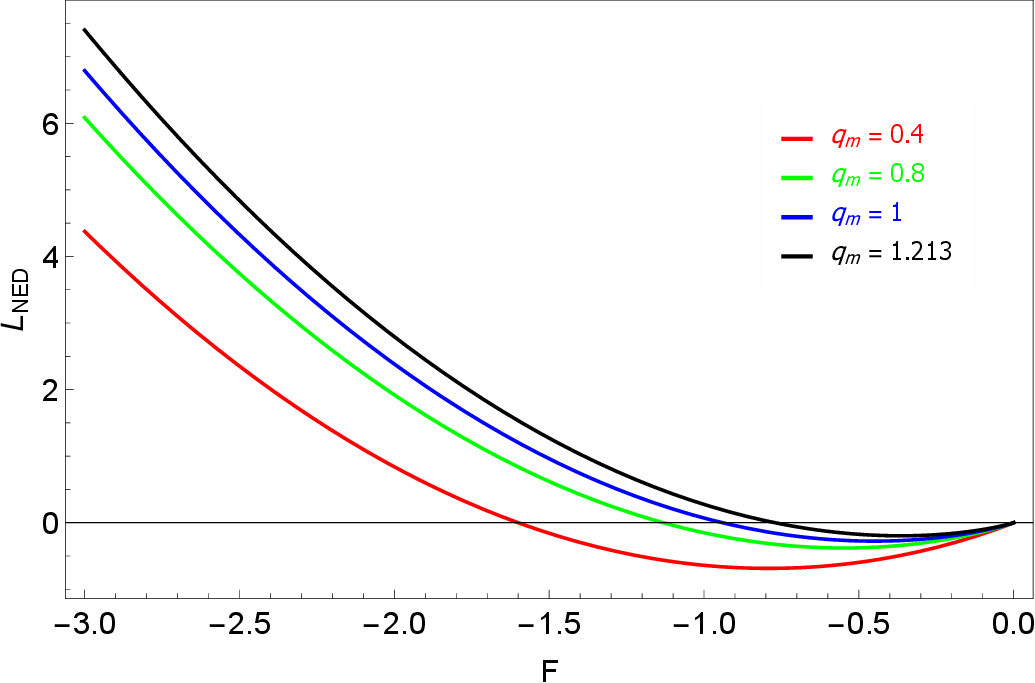}
    \caption{Plot of $L_{NED}$ as a function of $F$ with $\mathcal{C}=1,~M=1.$ Left: for fix $q_m=0.1$ and various values of $\beta$. Right: for fixed $\beta=\frac{1}{4}$ and various values of $q_m$}
    \label{fig:LNEDmassI}
\end{figure}

To obtain $L_{NED}$, we substitute \eqref{massI} into \eqref{Field_equation_1} and solve for $L_{NED}$. Thus, we obtain
\begin{align}
    L_{NED}(F) &= \mathcal{C}F^{1+1/4\beta} + \frac{2F}{\beta}Ei_x(y), \label{LNEDI}
\end{align}
where $\mathcal{C}$ is an integration constant and $Ei_x(y)$ is the exponential integral function where $x\equiv 1+1/\beta$ and $y \equiv -\frac{(-1)^{3/4}q_m^{3/2}F^{1/4}}{2^{3/4}M}$. Since $F$ is negative, this restricts the value of $\beta$ i.e., $1+\frac{1}{4\beta} = n$ where $n$ is an integer. In Fig~\ref{fig:LNEDmassI}, we illustrate the behaviour of $L_{NED}$ as a function of invariant $F$. These plots demonstrate clearly a modification of standard Maxwell Lagrangian. As can be seen from the plots, $L_{NED}$ approaches zero as $F\to 0$. With a given $\mathcal{C}$ and requiring that $L_{NED}$ should be a real value, one can show that $L_{NED}\sim F + O(F^{5/4})$ at small $F$. Interestingly, the no-go theorem states that the Einstein field equation couples to Lagrangian with the Maxwell behaviour at small $F$ (i.e., $L\to 0, L_F\to 1$ as $F\to 0$) does not admit static spherically symmetric purely electric solution with a regular centre \cite{Bronnikov:2000vy}. Let us remark that, our attempts to find regular black holes with electric charge are not successful. This is because with the purely electric gauge potential, the field equations reduces to a much more complicated second order differential equation comparing to \eqref{EFE00}.


\subsection{Fixed Lagrangian}\label{subsection:fixedL}
\begin{figure}[ht]
    \centering
    \includegraphics[width = 8.35cm]{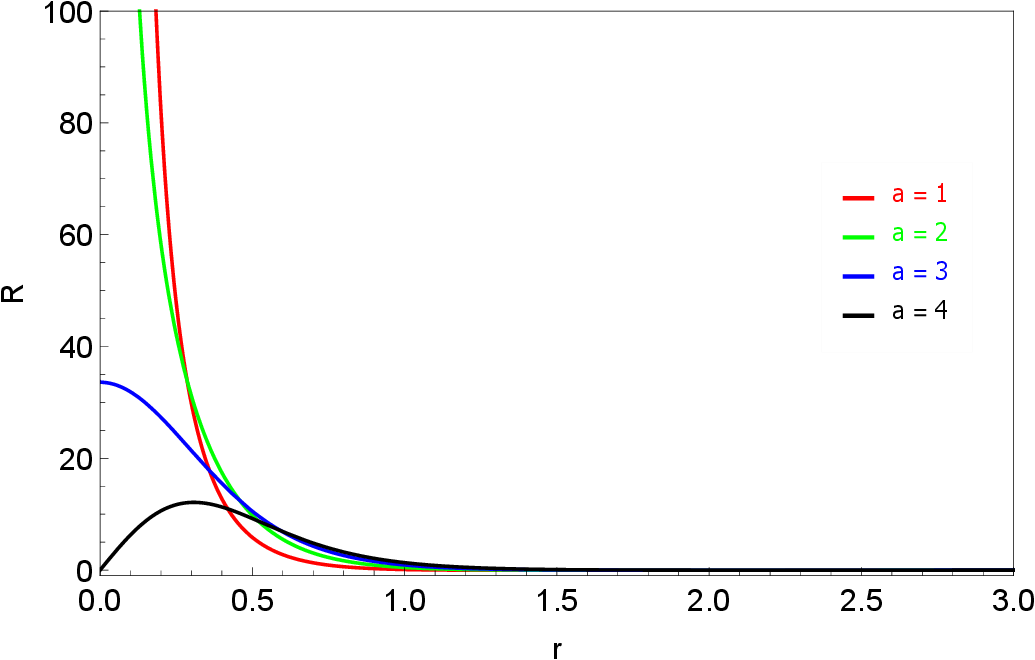}
    \includegraphics[width = 8cm]{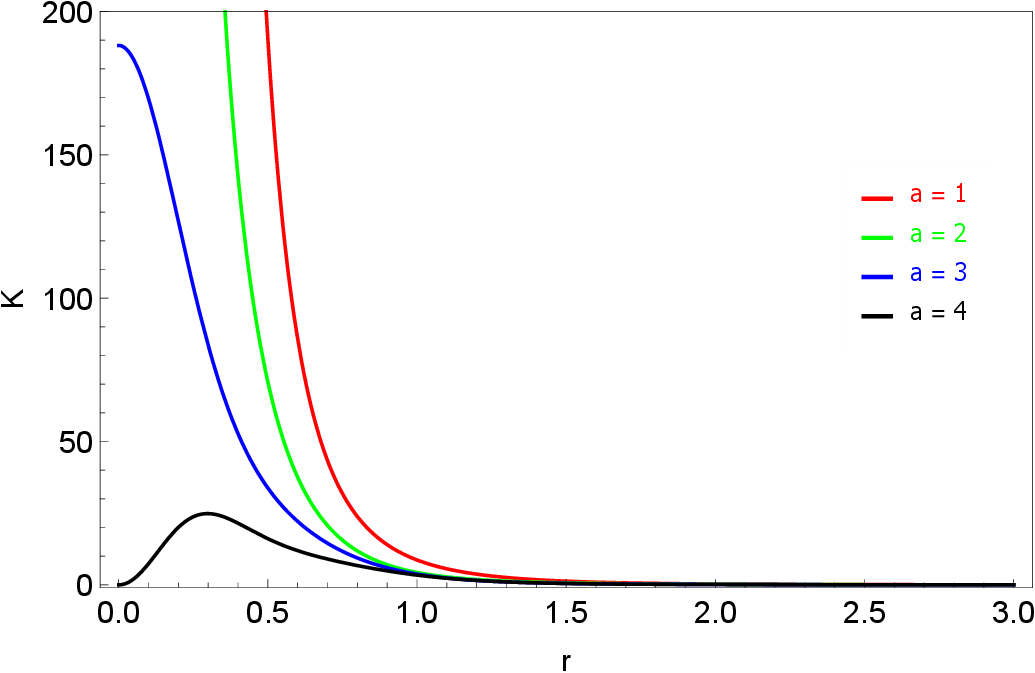}
    \caption{The Ricci scalar (Left) and the Kretchmann scalar (Right) as a function of $r$ with varying $a$ and $\beta=0.1,q=0.9,\alpha=1$ and $b=2$. }
    \label{fig:fixmass}
\end{figure}

Here, we solve \eqref{Field_equation_1} for the mass function when the matter Lagrangian is fixed. The Lagrangian of nonlinear electrodynamics is chosen to be
\begin{align}
    L_{NED}(F) &= -\frac{2a}{\alpha}\frac{\left(-4\alpha F\right)^{(b+3)/4}}{\left(1+\left(-4\alpha F\right)^{b/4}\right)^{1+a/b}}, \label{LNED}
\end{align}
where $a,b$ and $\alpha$ are positive-valued arbitrary constants. This Lagrangian is adopted from \cite{Fan:2016hvf,Lessa:2023xto} where the authors construct regular black holes in Einstein-NED and Einstein Cubic gravity, respectively. By inserting this Lagrangian into \eqref{Field_equation_1}, the following mass function is obtained
\begin{align}
    m(r)  &= M - \frac{q^3}{\alpha}(1+\beta) + \frac{q^3}{\alpha}\mathcal{Q}^{-a/b}(1+\beta) + \frac{a q^3}{\alpha}\mathcal{Q}^{-1-a/b}\left(\mathcal
{Q}-1\right)\beta, 
\end{align}
where $\mathcal{Q}(r) \equiv 1 + \left(\frac{q}{r}\right)^b$. $M$ is the gravitational mass and $q$ is an integration constant related to the magnetic charge $q_m=\frac{q^2}{\sqrt{2\alpha}}$. As it was pointed out in \cite{Toshmatov:2018cks},  we may define the effective mass $M_{eff}$ as the difference between gravitational mass $M$ and the magnetically induced mass $M_{em} = \frac{q^3}{\alpha}\left(1+\beta\right)$, i.e., $M_{eff} = M - M_{em}$. Then, regular black hole is obtained by letting $M = M_{em}$. Therefore, the metric function of regular black hole in $f(R,T)$ gravity coupled to nonlinear electrodyanmics source is 
\begin{align}
    A(r)  &=  1 - \frac{2q^3}{\alpha r}\mathcal{Q}^{-a/b}\left[(1+\beta) + a\beta\frac{\left(\mathcal{Q}-1\right)}{\mathcal{Q}}\right] \nonumber \\
    &= 1 - \frac{2q^3}{\alpha}\left(r^b + q^b\right)^{-a/b}\left[(1+\beta) + \frac{a\beta q^b}{r^b + q^b}\right]r^{a-1},  \label{GenA} 
\end{align}
From \eqref{GenA}, it appears that to avoid the singularity, one must take $a\geq 1$. But a closer investigation on the Ricci and the Kretechmann scalar reveals that to ensure the regularity of the solution as $r\to 0$, $a$ must be equal or greater than three ($a\geq 3$). This is demonstrated in Fig~\ref{fig:fixmass}. We observe that both scalar curvatures diverge as $r\to 0$ for $a<3$. The leading order term of both scalar curvatures are 
\begin{align}
    R &\sim r^{a-1},~~~~ K \sim r^{2a-6}.
\end{align}
This agrees with the results found in \cite{Fan:2016hvf}. For $a=4$, the maximum value of $R$ and $K$ are $12.11$ and $24.85$ for $r=0.3$ and $r=0.29$ respectively. For the remaining part of this work, we consider only the case where $a\geq 3$.

Now, we consider asymptotic structures of $A, R$ and $K$. As $r\to\infty$, we find that for $a\geq 3$
\begin{align}
    A &\sim 1 - \frac{2q^3}{\alpha r}(1+\beta) + O\left(\frac{1}{r^{b+1}}\right), \\
    R &\sim O\left(\frac{1}{r^5}\right),~~~~~~~~~~\text{for $b\leq 2$} , \\
    &\sim O\left(\frac{1}{r^{b+3}}\right),~~~~~~\text{for $b>2$}, \\
    K &\sim O\left(\frac{1}{r^6}\right).
\end{align}
The leading order of $A$ suggests that the solution \eqref{GenA} is asymptotically flat while the others display the regularity of the scalar curvatures at large $r$. The location of the black hole's event horizon is subtle without specifying $a$ and $b$. For the sake of demonstration, we consider three particular cases, i.e., (i) $a=3,b=2$, (ii) $a=3,b=3$ and (iii) $a=4,b=2$. The first two cases are chosen such that the Lagrangian \eqref{LNED} gives rise to the Bardeen-like and Hayward-like solutions \cite{Ayon-Beato:2000mjt,Hayward:2005gi,Fan:2016hvf}. The regular black holes for (i-iii) in $f(R,T)$ gravity are \\
(i) $a=3,b=2$ 
\begin{align}
    A_B(r) &= 1 - \frac{2q^3r^4}{\alpha\left(r^2 + q^2\right)^{5/2}}\left[(1+\beta) + \frac{q^2}{r^2}(1+4\beta)\right], \label{ansatz1}
\end{align}
(ii) $a=3,b=3$
\begin{align}
     A_H(r) &= 1 - \frac{2q^3r^5}{\alpha\left(r^3 + q^3\right)^{2}}\left[(1+\beta) + \frac{q^3}{r^3}(1+4\beta)\right],  \label{ansatz2}
\end{align}
(iii) $a=4,b=2$
\begin{align}
     A(r) &= 1 - \frac{2q^3r^5}{\alpha\left(r^2 + q^2\right)^{3}}\left[(1+\beta) + \frac{q^2}{r^2}(1+5\beta)\right].  \label{ansatz3}
\end{align}

\begin{figure}[ht]
    \centering
    \includegraphics[width = 5.4cm]{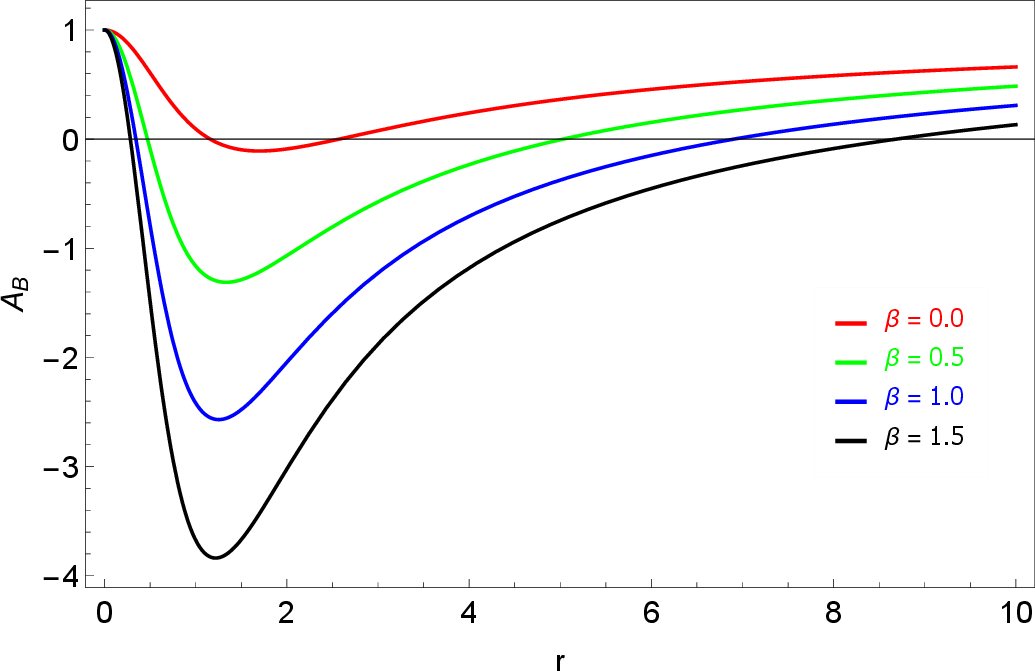}
    \includegraphics[width = 5.4cm]{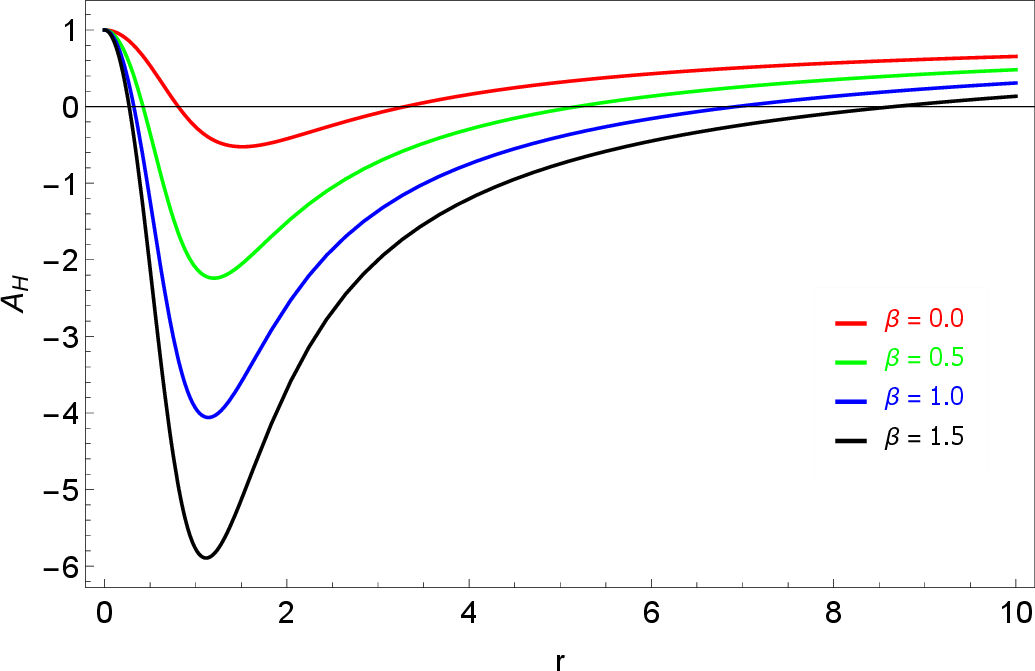}
    \includegraphics[width = 5.4cm]{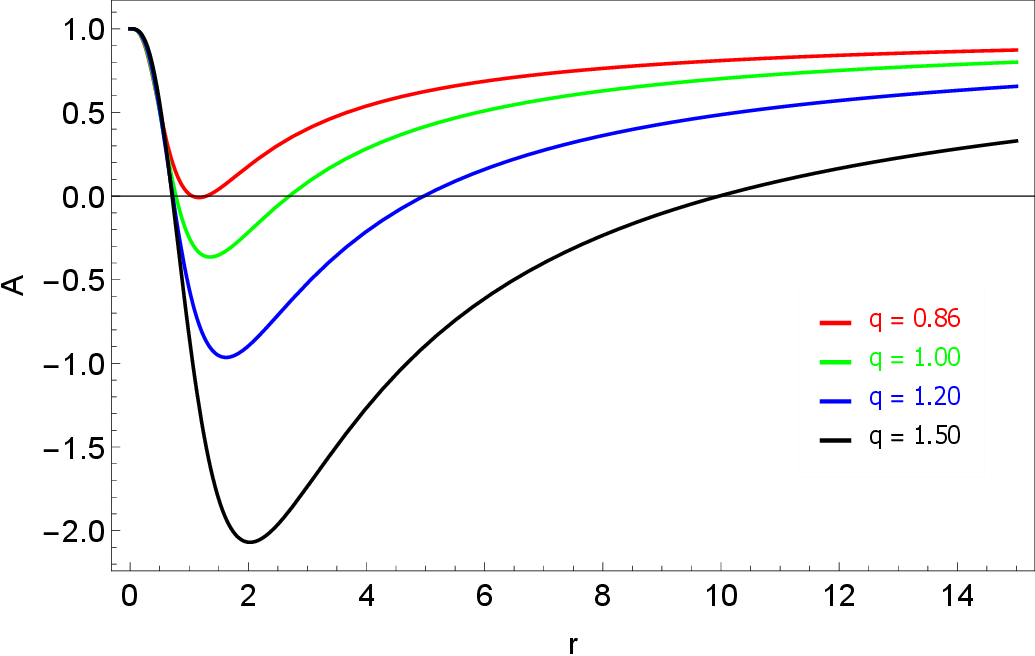}
    \caption{Example plots of $A(r)$ as a function of $r$ for $\alpha=1$. Left: the Bardeen-like solutions with $q=1.2$. Center: the Hayward-like solutions with $q=1.2$. Right: the regular black holes with $a=4,b=2,\beta=0.5$ for various values of $q$.}
    \label{fig:solexample}
\end{figure}
As $\beta\to 0$, the solutions $A_B$ and $A_H$ become regular black holes in general relativity i.e., the Bardeen and the Hayward solutions, respectively. The behaviours of these solutions are illustrated in Fig~\ref{fig:solexample}. The Bardeen-like and the Hayward-like solutions are shown for varying $\beta$. For the right figure, we fix $\beta=0.5$ and vary $q$ instead. It can be seen that extremal black holes and horizonless solution are also possible by varying $\beta$ or $q$. For example, in the third case, the two horizons coincide at $r=1.157$ with $q=0.85617015$. Moreover, it is clear that they are asymptotically flat as $A\to 1$ at large $r$.

\section{Energy conditions}\label{sec:energycond}
In this section, we consider null, weak and strong energy conditions (NEC, WEC, SEC) of the solutions discussed in the previous section. To consider the energy conditions in $f(R,T)$ gravity, let us re-write \eqref{EFE} as 
\begin{align}
    R_{\mu\nu} - \frac{1}{2}Rg_{\mu\nu} &= f^{-1}_R \left[ T_{\mu\nu} - f_T \left( T_{\mu\nu} + \Theta_{\mu\nu}\right)  - \left(g_{\mu\nu}\Box - \nabla_{\mu}\nabla_{\nu}\right)f_R    + \frac{1}{2}g_{\mu\nu}\left(f - R f_R\right)            \right], \nonumber \\
    &\equiv T^{(eff)}_{\mu\nu}.
\end{align}
where we have defined the effective energy-momentum tensor $T^{(eff)}_{\mu\nu}$. We identify ${T^{(eff)0}}_{0}=-\rho^{(eff)}$, ${T^{(eff)1}}_{1}=p^{(eff)}_{1}, {T^{(eff)2}}_{2}=p^{(eff)}_{2}$ and ${T^{(eff)3}}_{3}=p^{(eff)}_{3}$. The energy conditions in $f(R)$ gravity coupled to NED are discussed in \cite{Rodrigues:2015ayd}. In addition, the energy conditions of $f(R,T)$ gravity have been addressed properly in \cite{Alvarenga:2012bt,Sharif:2012ce,Rajabi:2021bdd}. These are 
\begin{align}
    NEC &: \rho^{(eff)} + p^{(eff)}_{1,2,3} \geq 0, \\
    WEC &: \rho^{(eff)} \geq 0, \rho^{(eff)} + p^{(eff)}_{1,2,3} \geq 0, \\
    SEC &: \rho^{(eff)} + p^{(eff)}_{1} + p^{(eff)}_{2} + p^{(eff)}_{3} \geq 0, \rho^{(eff)} + p^{(eff)}_{1,2,3} \geq 0.
\end{align}
To clarify the notation, $\rho^{(eff)} + p^{(eff)}_{1,2,3} \geq 0$ is simply $\rho^{(eff)} + p^{(eff)}_{i} \geq 0$ for $i=1,2,3$ separately. In this model, the non-vanishing diagonal components of the effective energy-momentum tensor are given explicitly by
\begin{align}
    \rho^{(eff)} &= -\left(\beta T + L_{NED}\right), \\
     p^{(eff)}_{1} &= \left(\beta T + L_{NED}\right), \\
     p^{(eff)}_{2} &= p^{(eff)}_{3} = L_{NED} + \frac{q_m^2}{r^4}L_F + \beta \left(T + \frac{2q_m^4}{r^8}L_{FF}\right).
\end{align}
Therefore, the energy conditions mentioned above reduce to
\begin{align}
    NEC &: \rho^{(eff)} + p^{(eff)}_{1,2} \geq 0, \label{NEC} \\
    WEC &: \rho^{(eff)} \geq 0, \rho^{(eff)} + p^{(eff)}_{1,2} \geq 0,  \label{WEC}\\
    SEC &: 2p^{(eff)}_{2}  \geq 0, \rho^{(eff)} + p^{(eff)}_{1,2} \geq 0, \label{SEC}.
\end{align}
Overall, we have four distinct inequalities. These will be considered in the following subsection for each regular black holes. 

\subsection{Energy conditions I}
Now, we consider the Lagrangian \eqref{LNEDI}. We also assume that $q_m$ and $M \geq 0$. It turns out that the energy conditions demand the following
\begin{align}
    NEC_2~\&~WEC_3~\&~SEC_3: \frac{e^{-q_m^2/2Mr}q_m}{r}\left(8Mr - q_m^2\right) \geq 0, \label{energy1}\\
    SEC_1: \frac{e^{-q_m^2/2Mr}q_m}{r}\left(4Mr - q_m^2\right) \geq 0. \label{energy2}
\end{align}
Note that, SEC$_1$ and SEC$_3$ refer to $2p^{(eff)}_{2}  \geq 0$ and $ \rho^{(eff)} + p^{(eff)}_{2} \geq 0$ respectively. The NEC$_1$, WEC$_1$, WEC$_2$ and SEC$_2$ are automatically satisfied. The NEC$_2$, WEC$_3$ and SEC$_3$ share similarities. Moreover, the SEC$_1$ provides another constraint on radial coordinate $r$ \eqref{energy2}. However, all the energy conditions are satisfied simultaneously in a region $r \geq \frac{q_m^2}{4M}$.

For parameter set chosen in Fig~\ref{fig:plotA}, we find that the NEC and the WEC are violated in the region $r < \{0.08, 0.125, 0.184\}$ for $q_m=0.8,1$ and $1.213$ respectively. However, these radii are much smaller than the inner event horizons. Thus, in these cases, the NEC and the WEC are satisfied in the exterior regions of the black holes. In contrast, the SEC$_1$ is found to be violated in a region between inner and outer horizon in $q_m=0.8$ and $1$ cases. While the SEC$_3$ \eqref{energy1} for these two cases requires $r \geq 0.08,0.125$, the inner horizon locates at $r=0.110$ and $r=0.232$ respectively. Therefore, the SEC$_3$ holds very well inside of the inner horizon. However, in the near extremal limit $q_m=1.213$, the SEC holds between two horizons. For these parameter sets, we find that all the energy conditions hold outside the outer event horizon. The NEC, WEC and SEC are not met in a region deep inside the black holes.

\subsection{Energy conditions II}
\begin{figure}[ht]
    \centering
    \includegraphics[width = 8cm]{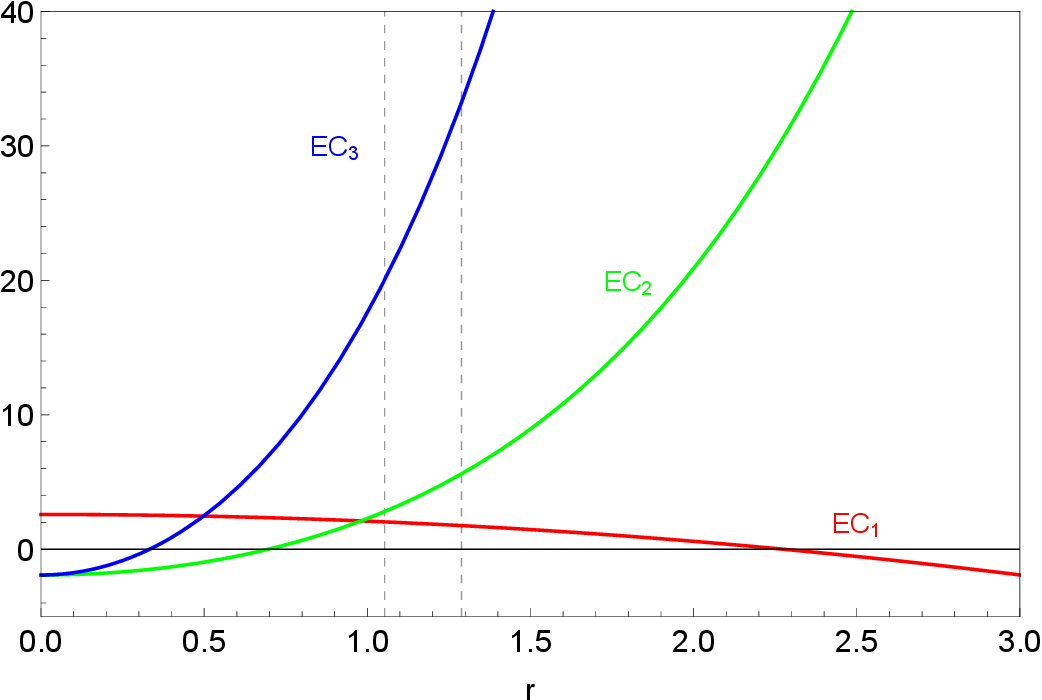}
    \includegraphics[width = 8cm]{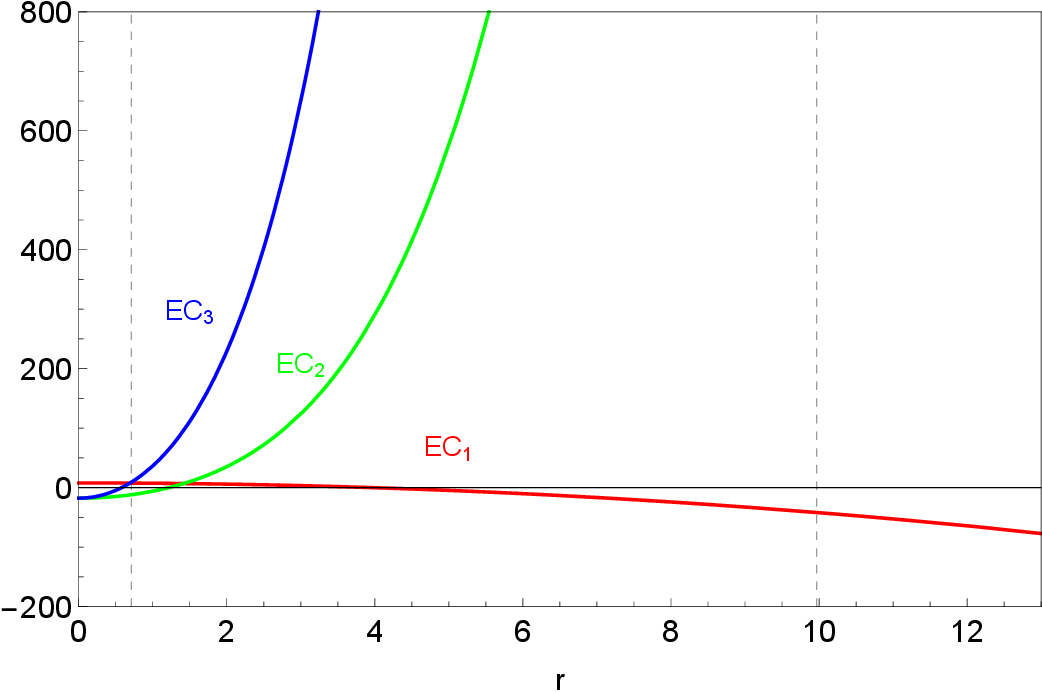}
    \caption{Plots of the energy conditions against $r$ for fixed $\beta=0.5$. The plot legends EC$_1$,EC$_2$,EC$_3$ correspond to the energy conditions \eqref{EC_1}--\eqref{EC_3}, respectively. Left: $q=0.86$, Right: $q=1.5$. The dashed vertical lines represent the location of inner and outer horizons.  }
   \label{fig:EC1}
\end{figure}

\begin{figure}[ht]
    \centering
    \includegraphics[width = 5.4cm]{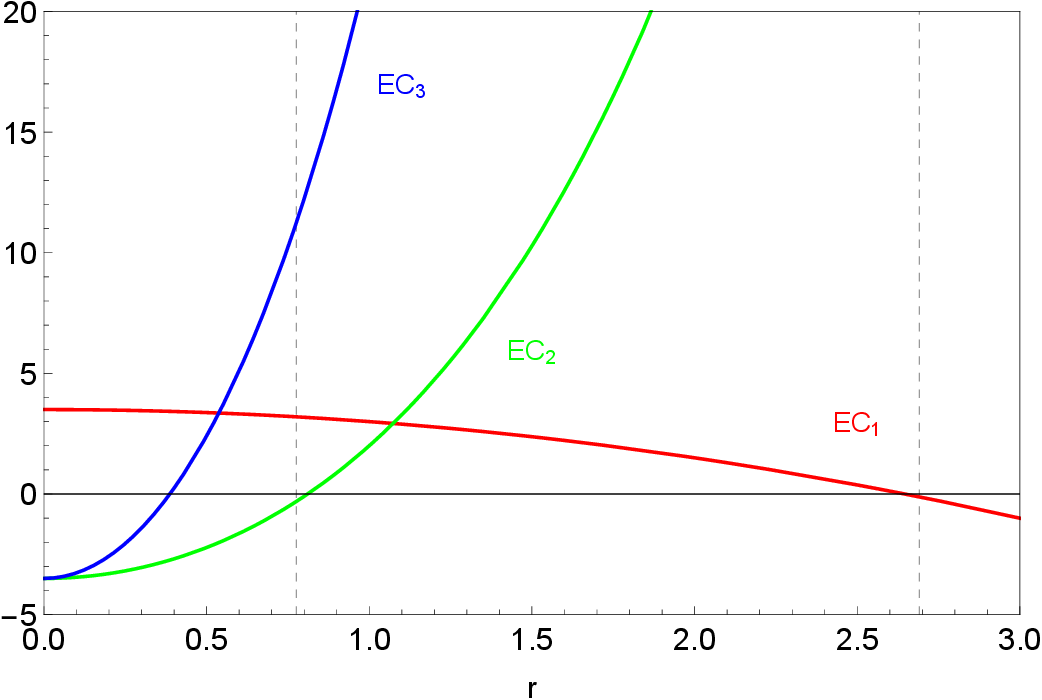}
    \includegraphics[width = 5.4cm]{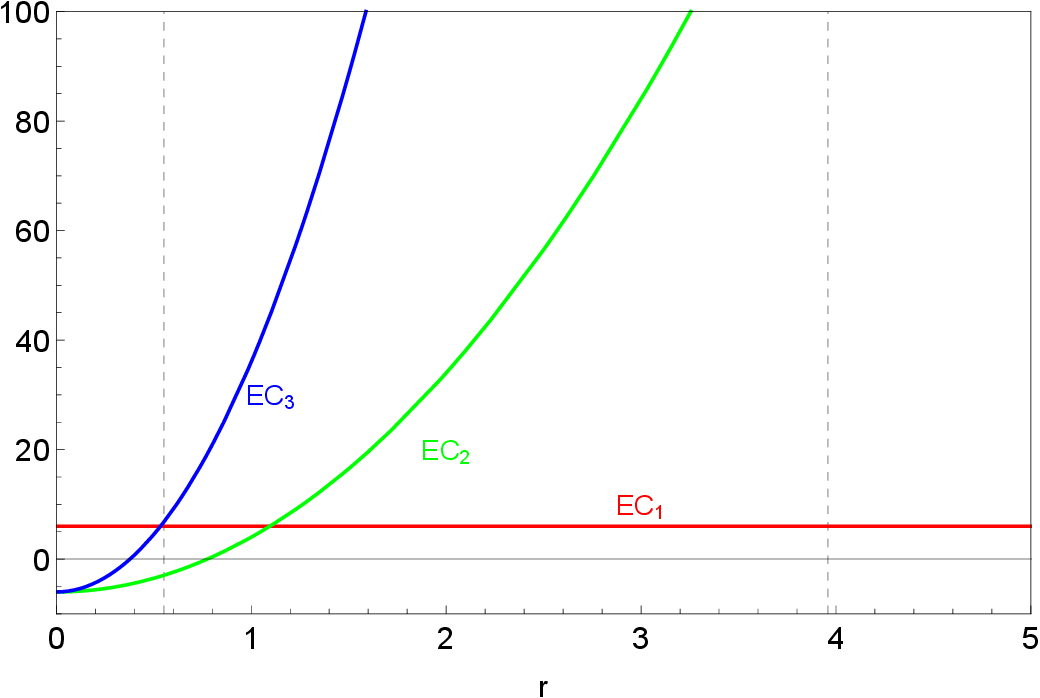}
    \includegraphics[width = 5.47cm]{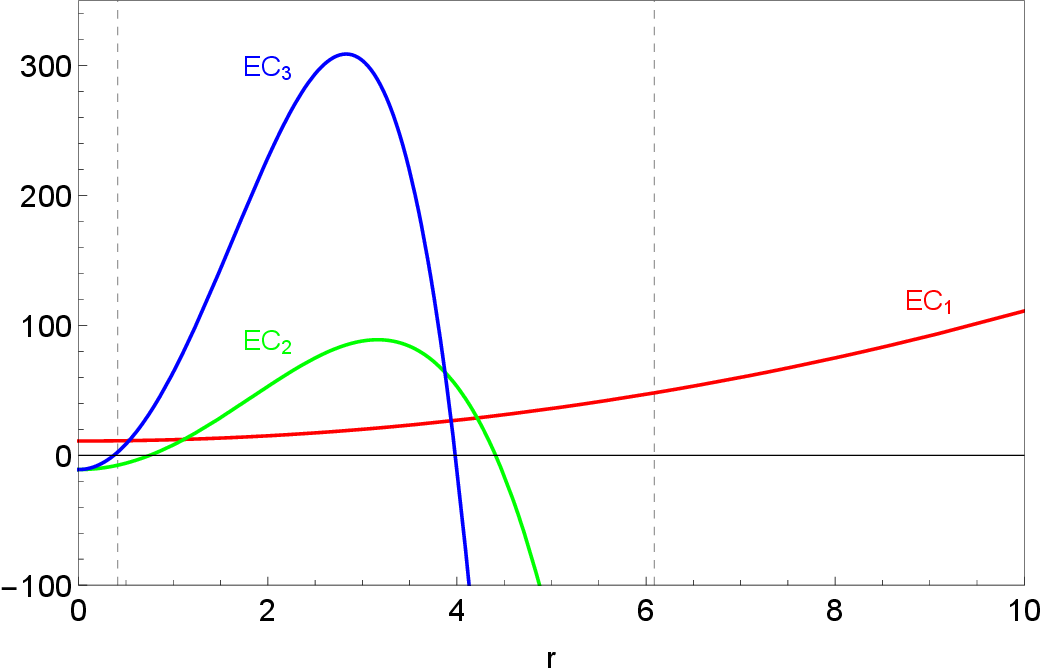}
    \caption{Plots of the energy conditions against $r$ for fixed $q=1$. The plot legends $EC_1,EC_2,EC_3$ correspond to the energy conditions \eqref{EC_1}--\eqref{EC_3}, respectively. Left: $\beta=0.5$, Centre: $\beta=1$, Right: $\beta=2$. The dashed vertical lines represent the location of inner and outer horizons.  }
  \label{fig:EC2}
\end{figure}

For the Lagrangian \eqref{LNED}, the energy conditions become complicated and lengthy without specifying $a$ and $b$. For this reason, we explicitly discuss the energy conditions for $a=4$ and $b=2$ case only. Similarly, the NEC$_2$, WEC$_3$ and SEC$_3$ are identical. The energy conditions are, therefore,
\begin{align}
    WEC_1: r^2 (\beta-1) + q^2 (1+5\beta) \geq 0, \label{EC_1} \\
    SEC_1: r^4 (1-\beta) + 10 q^2 r^2 \beta - q^4 (1+5\beta)  \geq 0, \label{EC_2} \\
    NEC_2~\&~WEC_3~\&~SEC_3: 5r^4(1-\beta) + 2 q^2 r^2 (2+19\beta) - q^4 (1+5\beta) \geq 0 \label{EC_3},
\end{align}
where we replace $q_m$ with $q$ to match with the notation used in subsection~\ref{subsection:fixedL}. The NEC$_1$, WEC$_2$ and SEC$_2$ are naturally satisfied. In Fig~\ref{fig:EC1} and \ref{fig:EC2}, we display the energy conditions as a function of $r$ for fixed $\beta$ and $q$, respectively. The energy conditions are violated if these curves become negative. When $q=0.86$, the WEC$_1$ holds continuously from the origin toward the exterior region. However, at certain radius outside of the black hole, the WEC$_{1}$ is violated (x-intercept is at $r=2.28$). In contrast, the other energy conditions hold from the certain radius inside the black hole all the way toward black hole's exterior. As we move away from near-extremal scenario, the EC$_1$ becomes negative relatively faster than the near-extremal case (x-intercept is at $r=3.97$). These are shown in Fig~\ref{fig:EC1}.

In addition, we explore how $\beta$ affects the energy conditions in Fig~\ref{fig:EC2}. The left figure shows similar behaviours of the energy conditions as mentioned in the previous figure. The WEC$_{1}$ is violated just prior to the outer event horizon (x-intercept is at $r=2.65$ while the outer horizon is at $r=2.69$). While the others are positive from the inner horizon. When $\beta=1$, the EC$_1$ becomes positive constant in $r$, therefore it is always satisfied. The other two remain positive right after the inner horizon. Remarkably, the energy conditions change dramatically for $\beta=2$. The EC$_2$ and EC$_3$ are positive for some particular region inside the black hole's outer horizon before rapidly become negative. The EC$_1$, on the other hand, holds throughout spatial coordinate $r$.

For this particular case i.e., $a=4$ and $b=2$, we find that the NEC and SEC are easily met at the black hole's exterior while the WEC will be violated at certain radius. However, appropriate selections of parameters can possibly make all the energy conditions satisfied.

\section{Quasinormal modes}\label{sec:QNMs}
A massive scalar field $(\Phi)$ on curved spacetime is described by the Klein-Gordon equation
\begin{align}
    \nabla_{\gamma}\nabla^{\gamma}\Phi - \mu^2 \Phi &= 0,
\end{align}
where $\mu$ is the scalar field's mass. In a spherical symmetric spacetime, the scalar field can be expressed as 
\begin{align}
    \Phi(t,r,\theta,\phi) &= \frac{R(r)}{r}e^{-i\omega t}Y(\theta,\phi),
\end{align}
where $Y(\theta,\phi)$ is the spherical harmonics. Under the spacetime metric \eqref{lineelement} (with $B=A^{-1}$), the Klein-Gordon equation takes the form
\begin{align}
    \frac{d^2R}{dr_{\ast}^2} + \left(\omega^2 - V(r)\right)R &= 0. \label{radialeq}
\end{align}
The effective potential is
\begin{align}
V(r) &= A(r)\left(\mu^2 + \frac{\ell(\ell+1)}{r^2} + \frac{A'(r)}{r}\right), \label{veff}
\end{align}
where $\ell$ is the spherical harmonic index. Moreover, we have introduced the tortoise coordinate defined by
\begin{align}
    r_{\ast} &= \int \frac{dr}{A(r)}.
\end{align}
The appropriate boundary conditions that lead to the quaisnormal mode are purely ingoing at the black hole's event horizon $r\to r_h$ or $r_{\ast}\to -\infty$ and no incoming flux at infinity $r,r_{\ast}\to\infty$. The frequencies $\omega$ corresponding to these boundary conditions will be discrete complex number or quasinormal frequencies. This complex frequency can be written in the form $\omega = \omega_R \pm i \omega_I$.

\begin{figure}[ht]
    \centering
    \includegraphics[width = 5.4cm]{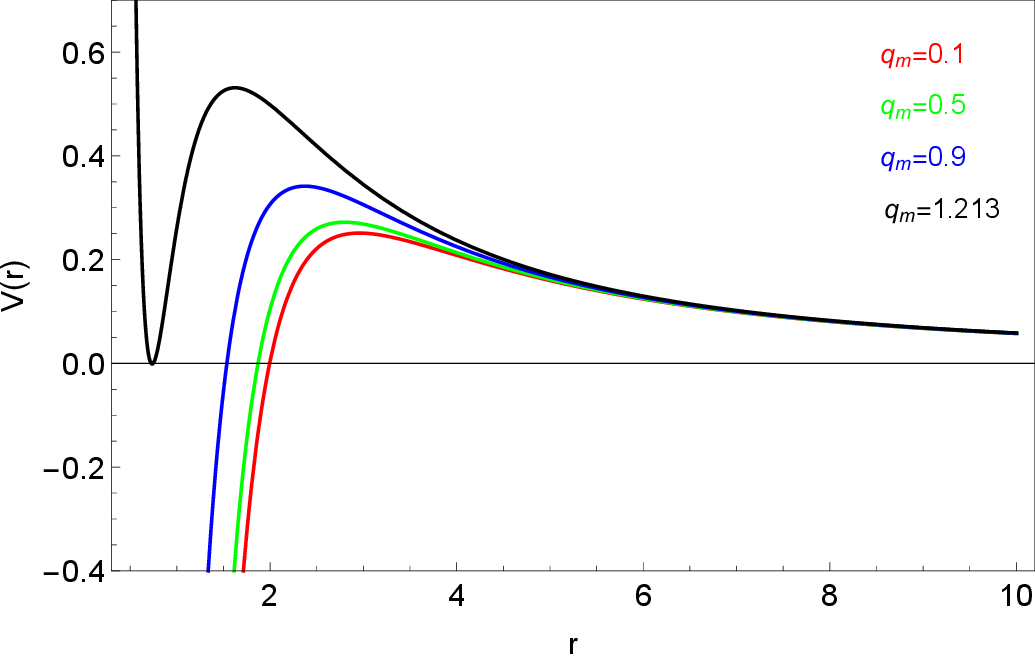}
    \includegraphics[width = 5.4cm]{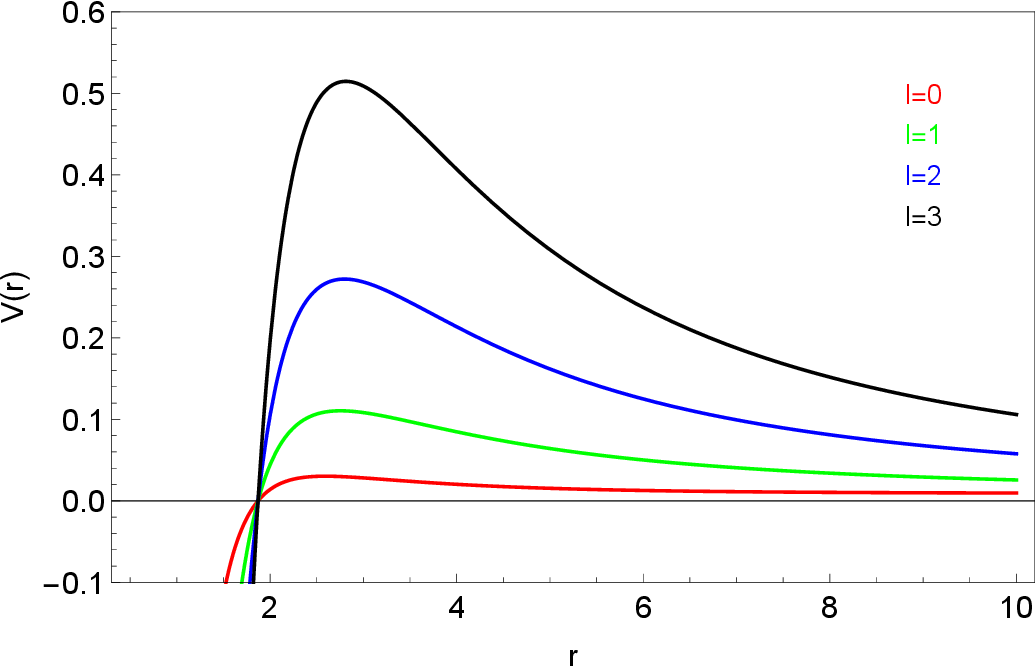}
    \includegraphics[width = 5.47cm]{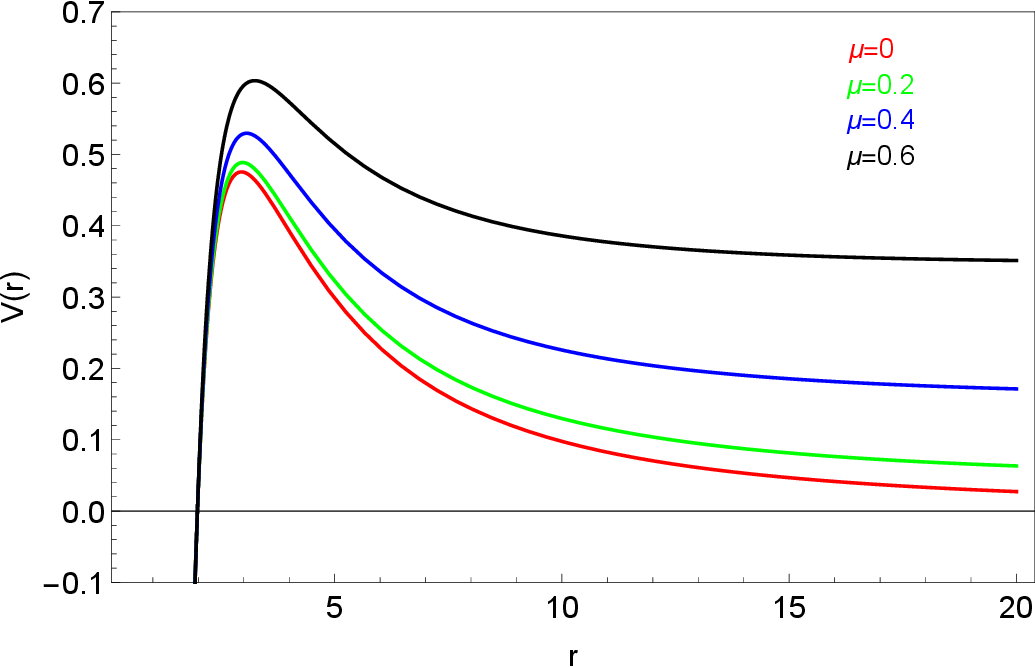}
    \caption{Plots of effective potential with mass function \eqref{massI} against $r$. In these plots, $M$ is fixed to $1$. Left: $\ell=2,\mu=0.1$ for various $q_m$, Centre: $q_m=0.5,\mu=0.1$  for various $\ell$, Right: $q_m=0.2,\ell=3$ for various $\mu$.}
  \label{fig:veff1}
\end{figure}

\begin{figure}[ht]
    \centering
    \includegraphics[width = 5.4cm]{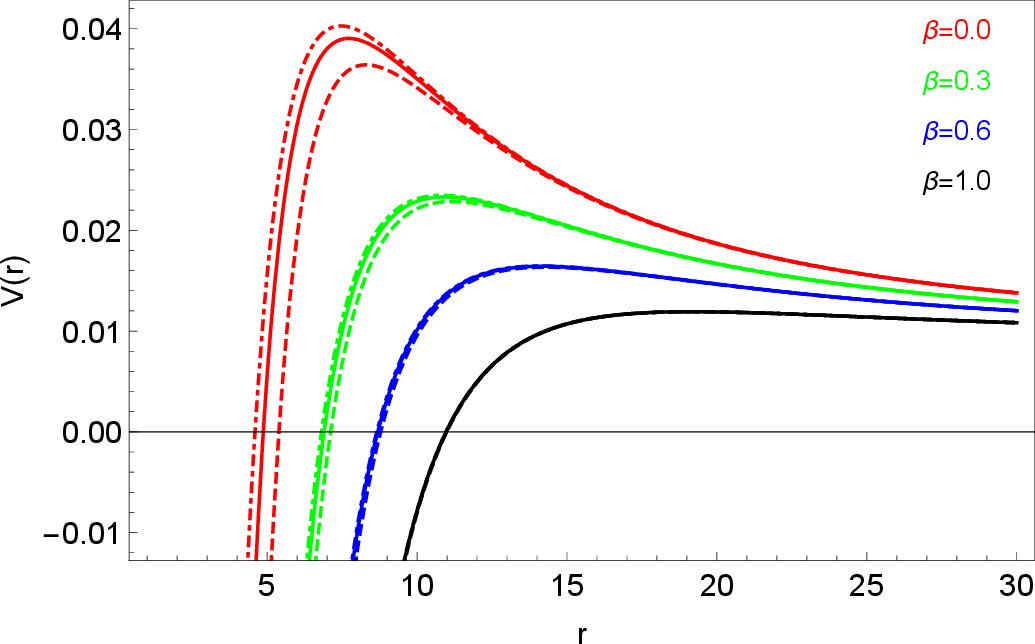}
    \includegraphics[width = 5.4cm]{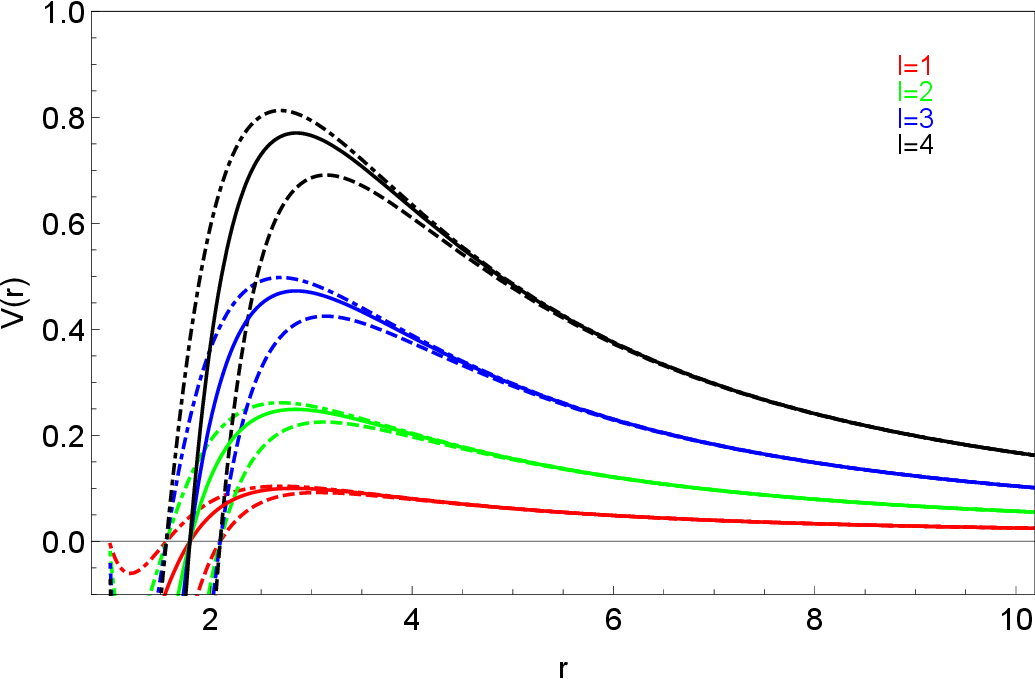}
    \includegraphics[width = 5.47cm]{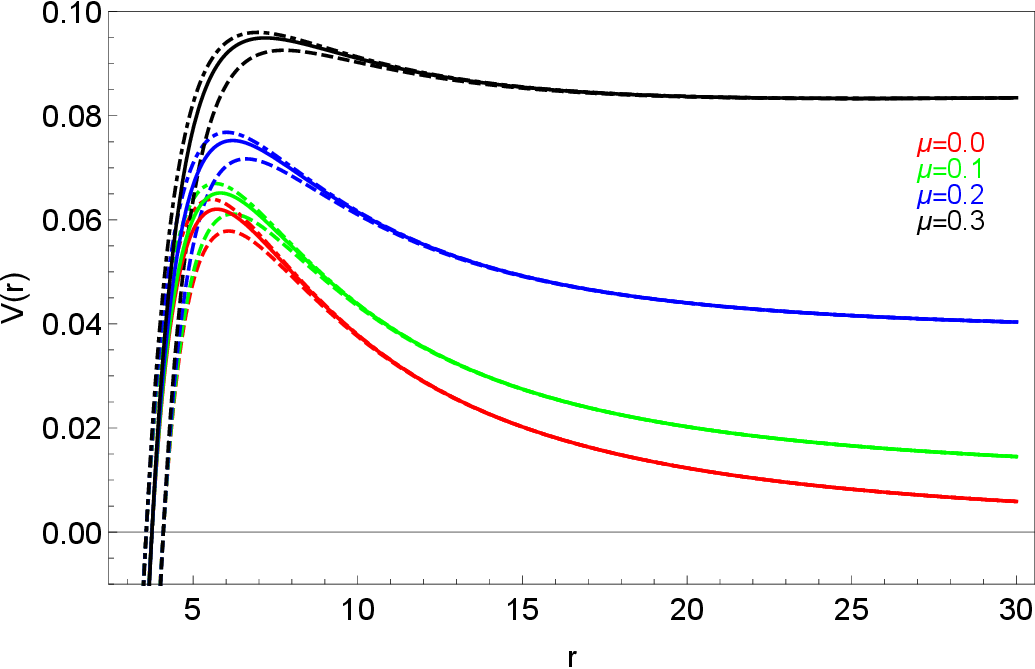}
    \caption{Plots of effective potential of three types of regular black holes against $r$. Left: fixed $\ell=2,\mu=0.1,\alpha=1,q=1.4$ for various $\beta$, Centre: fixed $\beta=0.45,\mu=0.1,\alpha=1,q=0.9$ for various $\ell$, Right: $\ell=2,\beta=0.2,\alpha=1,q=1.2$ for various $\mu$. The ansatz 1, 2 and 3 are illustrated by solid, dashed and dot-dashed lines.}
  \label{fig:veff2}
\end{figure}

Let's first consider the effective potential \eqref{veff} more explicitly. All the solutions considered in this work are asymptotically flat, therefore $V\to \mu^2$ as $r\to\infty$. Unless $A'(r) < 0$, the location where $V$ vanished is only determined by the roots of $A(r)$. The effective potential for several types of regular black holes are shown in Fig~\ref{fig:veff1}-\ref{fig:veff2}. For the mass function \eqref{massI}, the effective potential is illustrated in Fig~\ref{fig:veff1}. As the charge $q_m$ increases, the height of $V$ increases. We observe that the zeroth of $V$ occurs at the location of the black hole's outer horizon. More precisely, these potentials have another zeroth located at smaller $r$ which corresponds to the inner horizon. However, these are not explicitly displayed in the plots. In the extremal case (black solid line in the left figure), the potential possesses only one root. The central figure of Fig~\ref{fig:veff1} demonstrates the effect of harmonic index $\ell$ to the height of the effective potential. As $\ell$ decreases, the peak of $V$ decreases. The last figure illustrates that the peak of $V$ increases with scalar field's mass. Asymptotic value of $V$ approaches $\mu^2$ as expected. Remark that, similar plots are already explored in the case of scalar perturbations on the Bardeen solution \cite{Panotopoulos:2019qjk}.

Now, we consider the effective potential of the solutions reported in subsection \ref{subsection:fixedL}. We name the Bardeen-like, the Hayward-like and $a=4,b=2$ solution as ansatz 1, 2 and 3, respectively. The potentials are illustrated in Fig~\ref{fig:veff2}. In these plots, the ansatz 1, 2 and 3 are represented by solid, dashed and dot-dashed lines, respectively. The effects of $\ell$ and $\mu$ on the effective potential are qualitatively similar to the previous case as demonstrated in the central and right figures. We notice that the differences between the ansatz 1, 2 and 3 become more significant as $\ell$ or $\mu$ increases. In contrast, the differences are less apparent as $\beta$ increases. This is shown in the left figure. Moreover, the height of the potential decreases as $\beta$ increases. Remark that, in the case that $\beta=0$, the effective potential of the ansatz 1 (red solid line) is plotted in
\cite{Fernando:2012yw}.

\subsection{The Pad\'e averaged WKB approximation method}

To calculate the quasinormal frequencies, we employ the sixth order WKB approximation technique. With this method, it is possible to obtain the quasinormal frequencies $\omega$ via the following expression (up to sixth order) \cite{Konoplya:2003ii}
\begin{align}
\frac{i\left(\omega^2 - V_{max}\right)}{\sqrt{-2V''_{max}}} - \Lambda_2 - \Lambda_3 - \Lambda_4 - \Lambda_5 - \Lambda_6 &= n+\frac{1}{2}, \label{WKB}
\end{align}
where $V_{max}, V''_{max}$ are the effective potential and second derivative of the effective potential with respect to the tortoise coordinate evaluated at the maximum point of the potential. The overtone number is denoted by $n$. Iyer and Will find the correction terms up to the third order (hence $\Lambda_2,\Lambda_3$) \cite{PhysRevD.35.3621}. While later, Konoplya find three more correction terms $\Lambda_4,\Lambda_5$ and $\Lambda_6$ which are defined in \cite{Konoplya:2003ii}. To improve the numerical accuracy, the WKB approximation is extended to thirteen order including the Pad\'e averaged in \cite{Matyjasek:2017psv} where the quasinormal frequencies of the Schwarzschild and Reissner-Nordstr\"om black hole are reproduced. It turns out that with the Pad\'e averaged technique, many known results can be reproduced with great accuracy \cite{Matyjasek:2019eeu}. The Mathematica code for calculating quasinormal frequencies up to thirteen order WKB with improved Pad\'e averaged is provided in \cite{Konoplya:2019hlu}. Thus, we will implement the code for computing quasinormal frequencies in this work. 
We remark that, throughout this section, the parameter $\alpha$ is substituted by $\frac{q^3}{M}(1+\beta)$. 

\begin{widetext}
\begin{table}[htb]
{\centering
\begin{center}
\caption{The quasinormal frequencies computed by sixth order WKB with the Pad\'e averaged for $M=1$. Upper: the $\omega$ for solution \eqref{massI} for $\ell=1$ and $\ell=2$ (in parentheses) with fixed $\mu=0.1$. Lower: the $\omega$ for ansatz 1 for $\beta=0,q=0.76$  and $\ell=2$. Here, we only display $n=0$ mode.}
\vspace{0.3cm}
\setlength{\tabcolsep}{7pt}
\begin{tabular}{cccc}
\hline\\[-10pt]
$q_m$ & Pad\'e averaged WKB & Error estimation & Results from \cite{Panotopoulos:2019qjk} \\[2pt]  \hline
\\[-10pt] 

 $0.2$ & $0.299367 - 0.095199 i$ & $1.534997 \times 10^{-7}$ & $0.2993 - 0.0953i$ \\
       & $(0.490046 - 0.095900 i)$ & $(2.466144 \times 10^{-7})$ & $(0.4900 - 0.0959i)$ \\ 
 $0.4$ & $0.305515 - 0.095912 i$ & $9.141781 \times 10^{-8}$ & $0.3055 - 0.0960i$ \\
       & $(0.500232 - 0.096563 i)$ & $(2.412731 \times 10^{-7)}$ & $(0.5002 - 0.0966i)$ \\
 $0.6$ & $0.316811 - 0.097033 i$ & $2.341719 \times 10^{-6}$ & $0.3168 - 0.0971i$ \\
       & $(0.518964 - 0.097606 i)$ & $(2.100739 \times 10^{-7})$ & $(0.5190 - 0.0976i)$ \\
 $0.8$ & $0.335444 - 0.098326 i$ & $7.426319 \times 10^{-6}$ & $0.3354 - 0.0984i$ \\
       & $(0.549924 - 0.098795 i)$ & $(1.141331 \times 10^{-7})$ & $(0.5499 - 0.0988i)$ \\
\hline  
\hline\\[-10pt]
$\mu$ & Pad\'e averaged WKB & Error estimation & Results from \cite{Fernando:2012yw} \\[2pt]  \hline
\\[-10pt] 

 $0.1$ & $0.553916 - 0.079097 i$ & $2.211318 \times 10^{-7}$ & $0.553910 - 0.0791170 i$ \\
      
 $0.2$ & $0.560827 - 0.077603 i$ & $9.480210 \times 10^{-7}$ & $0.560821 - 0.0776240 i$ \\
       
 $0.3$ & $0.572506 - 0.074963 i$ & $3.869737 \times 10^{-7}$ & $0.572499 - 0.0749856 i$ \\
      
 $0.4$ & $0.589209 - 0.070920 i$ & $5.633078 \times 10^{-7}$ & $0.589201 - 0.0709443i$ \\

 $0.5$ & $0.611315 - 0.065028 i$ & $1.348216 \times 10^{-6}$ & $0.611305 - 0.0650541i$ \\
      
\hline  
\end{tabular}
\label{Tab:1}
\end{center}}
\end{table}
\end{widetext}

As a consistency check, we list the $n=0$ quasinormal frequencies in Table~\ref{Tab:1}. In this table, we reproduce the results already obtained in refs \cite{Panotopoulos:2019qjk,Fernando:2012yw} which are shown in the rightmost column. We implement the sixth order WKB with Pad\'e average to compute the quasinormal frequencies of massive scalar perturbation with spacetime background given by the mass function \eqref{massI} (the upper table) and ansatz 1 with $\beta=0$ (the lower table). The error estimation denotes the root mean square error corresponding to the sixth order WKB with Pad\'e approximation. The black hole's mass is set to unity. The upper table displays $\omega$ as a function of $q_m$ for $\ell=1$ and $\ell=2$ (in parentheses). Both real and imaginary parts increase as the black hole's charge increases (in magnitude). The lower table investigates the effect of scalar field's mass $\mu$ on the $ \omega$ with fixed $q=0.76$ and $\ell=2$. The real part of $\omega$ increases with $\mu$ whereas the imaginary part decreases with $\mu$. Apparently, the sixth order Pad\'e averaged WKB approximation method agrees with those results found earlier.  

Now, we shall turn our attention to the QNMs of regular black holes of ansatz 1, 2 and 3, \eqref{ansatz1}--\eqref{ansatz3}. In Table~\ref{Tab:2}--\ref{Tab:4}, the quasinormal frequencies with $\ell=0-2$  are displayed as a function of $q$ for Bardeen-like, Hayward-like and $a=4,b=2$ solutions respectively. For comparison, in these tables, we fix $M=1,\beta=0.1$ and scalar field's mass is $0.1$. Since the WKB approximation works very well when $\ell>n$ \cite{Panotopoulos:2019qjk,Cardoso:2003vt}, we consider $\ell=1,n=0$, $\ell=2,n=0$ and $n=1$ cases. Despite the $\ell=n=0$ case might not be well-approximated by the WKB method, we will include them in the tables since they are the most fundamental modes. It turns out that the quasinormal frequencies of the ansatz 1--3 share similar trends. As the black hole's charge $q$ increases, the real part of $\omega$ increases while the imaginary part becomes less negative. Various studies on QNMs of regular black holes also report the similar trend \cite{Fernando:2012yw,Wahlang:2017zvk,Panotopoulos:2019qjk,Liu:2020lwc}. With increasing angular index $\ell$, the real part increases. In contrast, the effect of $\ell$ on $\omega_I$ is non-trivial. At first, the imaginary part decreases (in magnitude) when $\ell$ moves from zero to one. Later, the imaginary part increases once again as $\ell=2$. Lastly, both $\omega_R$ and $\omega_I$ decrease as the overtone number $n$ is larger. We observe that the quasinormal frequencies of these regular black holes (ansatz 1--3) marginally differ from each other. This should not be surprised because the effective potentials of these ansatzes (Fig~\ref{fig:veff2}) are nearly identical. Therefore, for the remaining part of this article, we will particularly focus only on the ansatz 3 for the sake of presentation.


       
      
 

\begin{widetext}
\begin{table}[htb]
{\centering
\begin{center}
\caption{The Bardeen-like solution with $M=1,\beta=0.1$ and $\mu=0.1$.}
\vspace{0.3cm}
\setlength{\tabcolsep}{7pt}
\begin{tabular}{cccccc}
\hline\\[-10pt]
$\ell$ &  $n$  & $q=0.1$ & $q=0.3$ & $q=0.5$ & $q=0.7$ \\[2pt]  \hline
\\[-10pt] 

 $0$ & $0$ & $0.113977 - 0.096346 i$ & $0.115682 - 0.095390 i$ & $0.119425 - 0.093134 i$ & $0.123845 - 0.086839 i$ \\
       
 $1$ & $0$ & $0.297824 - 0.094875  i$ & $0.301256 - 0.094145 i$ & $0.308793 - 0.092224 i$ & $0.322230 - 0.087376 i$ \\
      
 $2$ & $0$ & $0.487471 - 0.095589  i$ & $0.493002 - 0.094832 i$ & $0.505246 - 0.092863 i$ & $0.527654 - 0.087942 i$ \\
     & $1$ & $0.466262 - 0.292988 i$ & $0.472838 - 0.290381 i$ & $0.487197 - 0.283678 i$ & $0.512242 - 0.267288 i$ \\
 
\hline  
\end{tabular}
\label{Tab:2}
\end{center}}
\end{table}
\end{widetext}


       
      
 

\begin{widetext}
\begin{table}[htb]
{\centering
\begin{center}
\caption{The Hayward-like solution with $M=1,\beta=0.1$ and $\mu=0.1$.}
\vspace{0.3cm}
\setlength{\tabcolsep}{7pt}
\begin{tabular}{cccccc}
\hline\\[-10pt]
$\ell$ &  $n$  & $q=0.1$ & $q=0.3$ & $q=0.5$ & $q=0.7$ \\[2pt]  \hline
\\[-10pt] 

 $0$ & $0$ & $0.113739 - 0.096406 i$ & $0.113850 - 0.096230 i$ & $0.114325 - 0.095620 i$ & $0.115884 - 0.094445 i$ \\
       
 $1$ & $0$ & $0.297418 - 0.094953 i$ & $0.297639 - 0.094813 i$ & $0.298484 - 0.094270 i$ & $0.300409 - 0.092945 i$ \\
      
 $2$ & $0$ & $0.486816 - 0.095670 i$ & $0.487162 - 0.095532  i$ & $0.488481 - 0.094994 i$ & $0.491519 - 0.093689 i$ \\
     & $1$ & $0.465483 - 0.293268 i$ & $0.465938 - 0.292810 i$ & $0.467658 - 0.291022 i$ & $0.471516 - 0.286672 i$ \\
 
\hline  
\end{tabular}
\label{Tab:3}
\end{center}}
\end{table}
\end{widetext}


       
      
 

\begin{widetext}
\begin{table}[htb]
{\centering
\begin{center}
\caption{The ansatz 3 ($a=4,b=2$) with $M=1,\beta=0.1$ and $\mu=0.1$.}
\vspace{0.3cm}
\setlength{\tabcolsep}{7pt}
\begin{tabular}{cccccc}
\hline\\[-10pt]
$\ell$ &  $n$  & $q=0.1$ & $q=0.3$ & $q=0.5$ & $q=0.7$ \\[2pt]  \hline
\\[-10pt] 

 $0$ & $0$ & $0.114032 - 0.096285 i$ & $0.116337 - 0.095010 i$ & $0.121850 - 0.091256 i$ & $0.122825 - 0.080233  i$ \\
       
 $1$ & $0$ & $0.297963 - 0.094847i$ & $0.302605 - 0.093826 i$ & $0.313203 - 0.090768 i$ & $0.333047 - 0.080207  i$ \\
      
 $2$ & $0$ & $0.487694 - 0.095560 i$ & $0.495182 - 0.094504 i$ & $0.512499 - 0.091388 i$ & $0.547741 - 0.080482 i$ \\
     & $1$ & $0.466529 - 0.292887  i$ & $0.475418 - 0.289264 i$ & $0.495503 - 0.278719 i$ & $0.529632 - 0.243894 i$ \\
 
\hline  
\end{tabular}
\label{Tab:4}
\end{center}}
\end{table}
\end{widetext}

\begin{figure}[ht]
    \centering
    \includegraphics[width = 8cm]{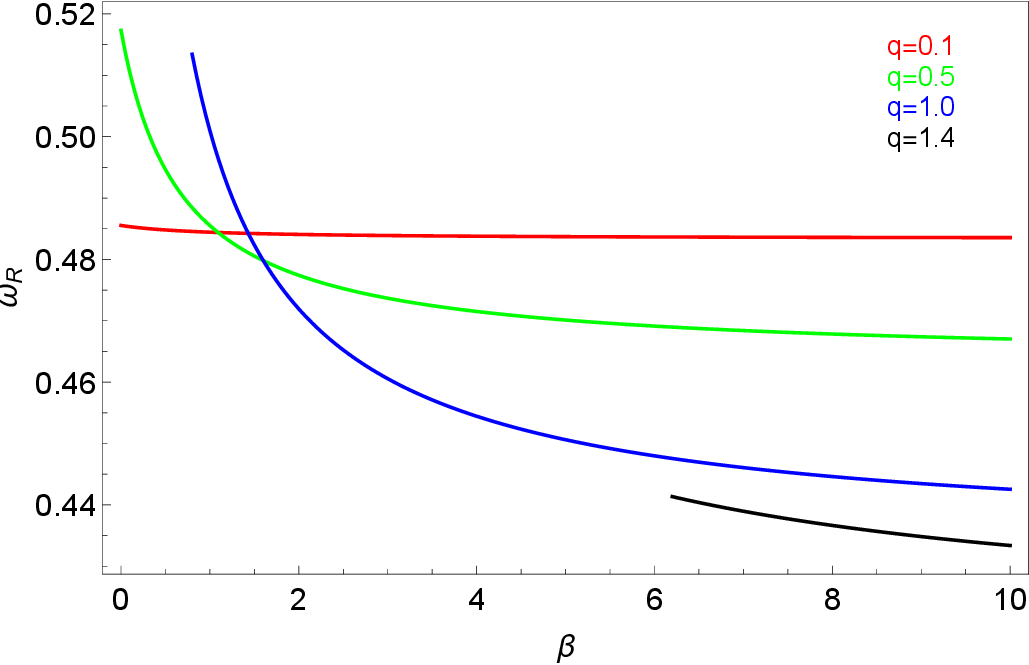}
    \includegraphics[width = 8.2cm]{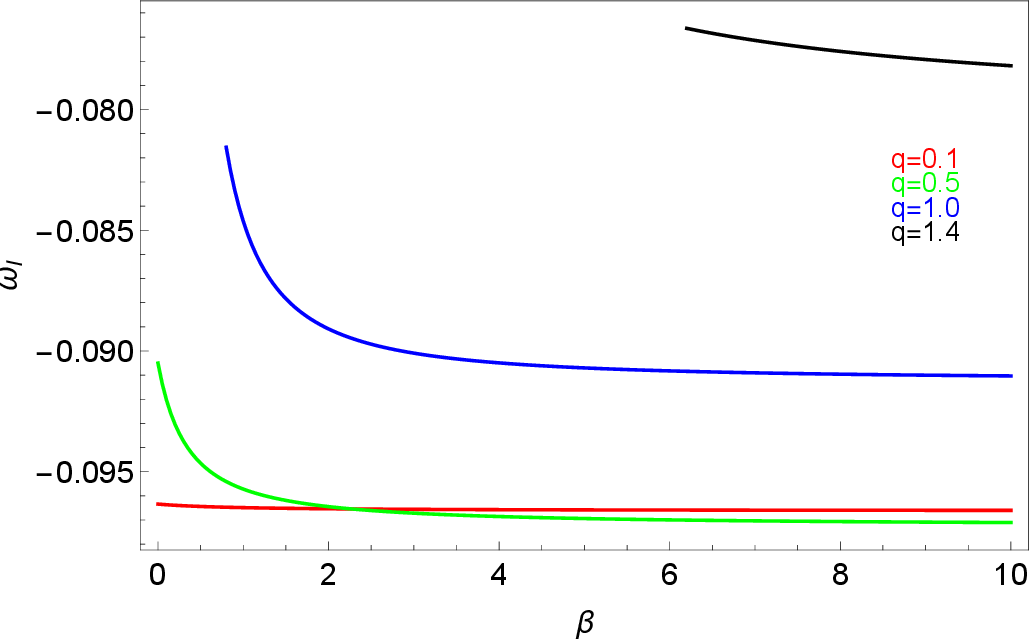}
    \caption{The real part and imaginary part of quasinormal frequencies of regular black holes (ansatz 3) as function of $\beta$ for various values of $q$. With $M=1,\ell=2,n=0$ and $\mu=0.05$.}
  \label{fig:qnm1}
\end{figure}

\begin{figure}[ht]
    \centering
    \includegraphics[width = 8cm]{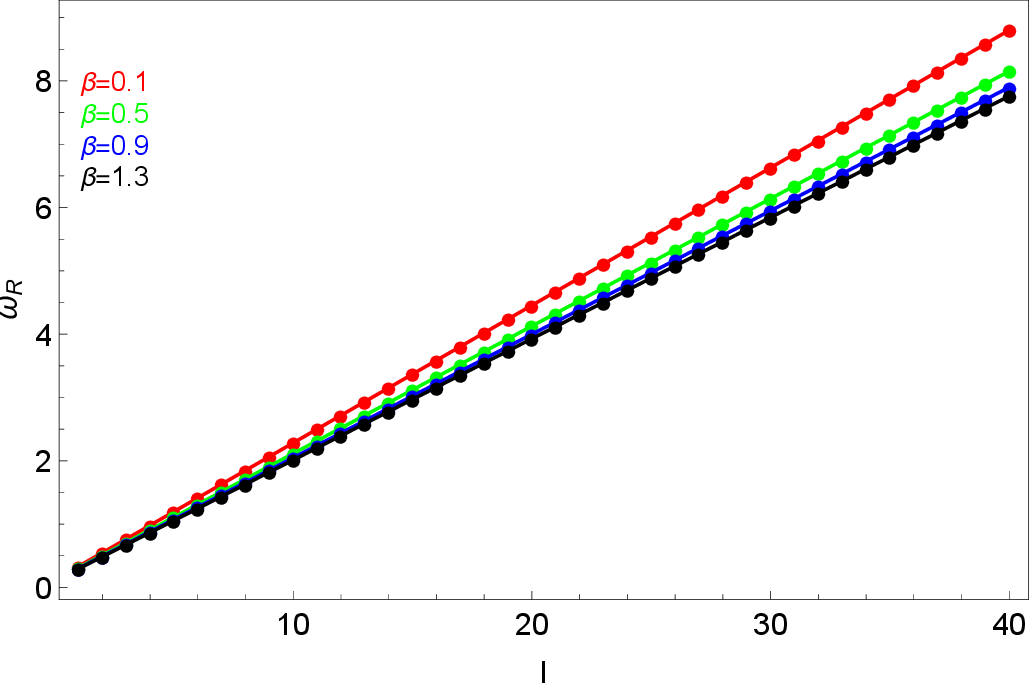}
    \includegraphics[width = 8.2cm]{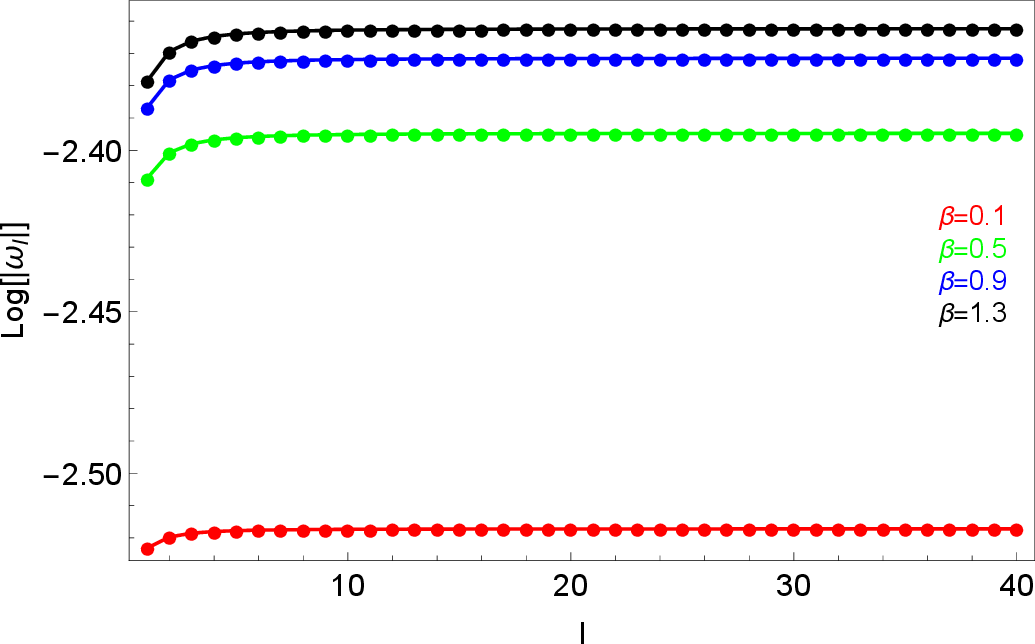}
    \caption{The real part and imaginary part of quasinormal frequencies of regular black holes (ansatz 3) as function of $\ell$ for various values of $\beta$. With $M=1,q=0.7,n=0$ and $\mu=0.1$. For the imaginary part, the $\ln (|\omega_I|)$ is plotted instead of $\omega_I$.}
  \label{fig:qnm2}
\end{figure}

\begin{figure}[ht]
    \centering
    \includegraphics[width = 8cm]{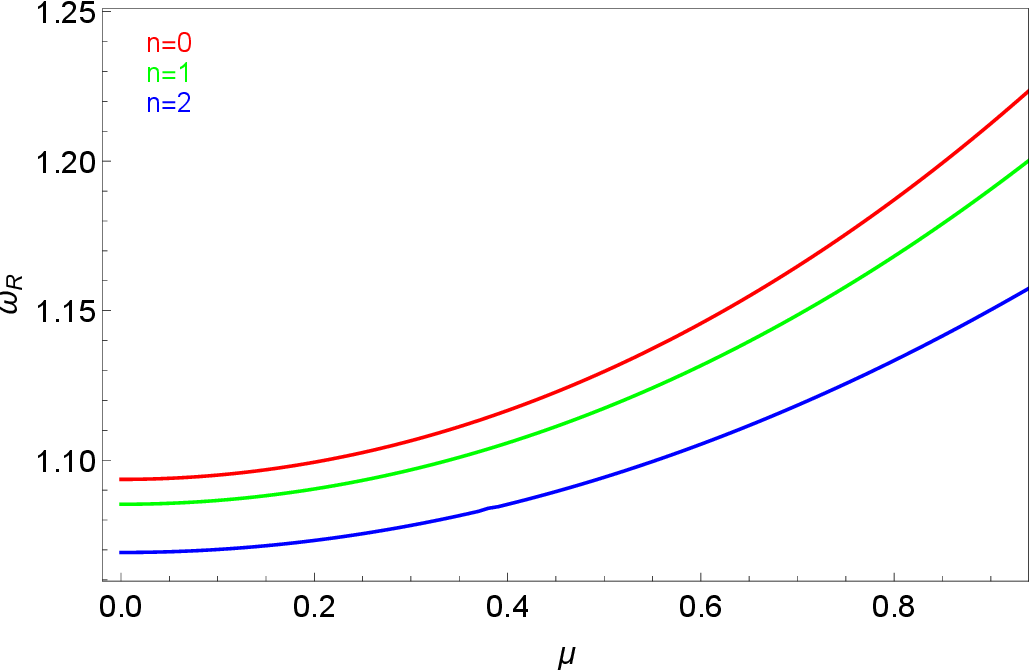}
    \includegraphics[width = 8.2cm]{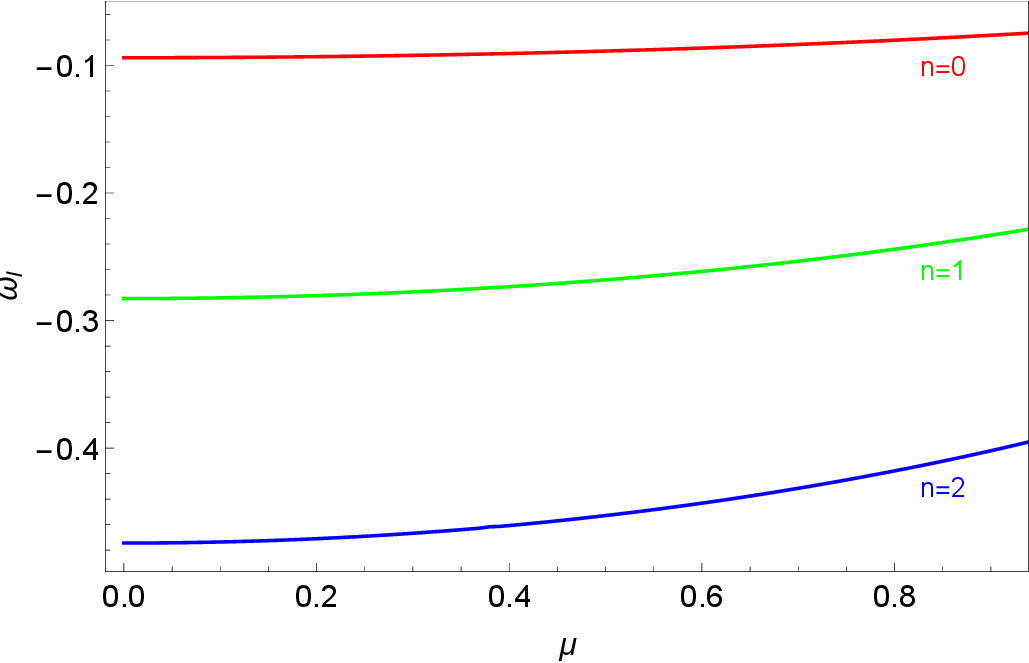}
    \caption{The real part and imaginary part of quasinormal frequencies of regular black holes (ansatz 3) as function of $\mu$ for various values of $n$. With $M=1,\beta=0.1,q=0.4$ and $\ell=5$.}
  \label{fig:qnm3}
\end{figure}

We explore how the coupling constant $\beta$ affects the quasinormal frequencies $\omega$ in Fig~\ref{fig:qnm1}. In this plot, we choose four particular values of black hole's charge $q=0.1,0.5,1.0$ and $1.4$. As $\beta$ increases, the real part and imaginary part of $\omega$ decrease. The decrease of frequencies becomes less obvious at larger $\beta$. In addition, the change in $\omega$ can be clearly seen as $q$ increases. As $q$ gets larger, both $\omega_R$ and $\omega_I$ are smaller (in magnitude). It is worth mentioning that at $q=1.0$ and $q=1.4$, there are no regular black holes for $\beta < 0.76$ and $\beta<6.2$ respectively.  

In Fig~\ref{fig:qnm2}, the dependence of $\omega$ on spherical harmonic index $\ell$ is illustrated. The real part of $\omega$ increases monotonically with $\ell$. In addition, the differences in $\omega_R$ between each fixed $\beta$ become more evident at large $\ell$. To demonstrate the change in $\omega_I$, we express them with $\ln (|\omega_I|)$. At small $\ell$, $\omega_I$ varies drastically with $\ell$, but as $\ell$ gets bigger, the change in $\omega_I$ becomes less significant. Remark that, 
the imaginary parts of quasinormal frequencies are more negative as $\ell$ increases. These trends are also observed for QNMs of regular black holes \cite{Panotopoulos:2019qjk}, Bardeen black holes \cite{Fernando:2012yw} and Bardeen de-Sitter black holes \cite{Wahlang:2017zvk}. It can be seen from the plots that $\omega_R$ and $\omega_I$ decrease as $\beta$ increases which are in agreement with what was discussed in the previous figure.  

In Fig~\ref{fig:qnm3}, we demonstrate the effect of scalar field's mass $\mu$ on the quasinormal frequencies. As the field's mass increases, the real part of the frequencies increases, while the imaginary part becomes smaller as $\mu$ increases. The lowest overtone mode ($n=0$) has larger $\omega_R$ and $\omega_I$ comparing to the higher overtone modes. We notice that $\omega_R$ increases monotonically along with the $\mu$. In contrast, for the first overtone ($n=1$) of QNMs of Bardeen black hole \cite{Fernando:2012yw}, its real part approaches a certain maximum value and then decreases with the scalar field mass. 
In addition, a study on massive scalar perturbation on Reissner-Nordstr\"om black hole reveals that it is possible to have arbitrarily long live modes or \textit{quasi-resonance} modes as the scalar field's mass increases \cite{Ohashi:2004wr,Konoplya:2004wg}. Similarly, our results are expected to respect this behaviour. However, the WKB approximation method is not sufficient to accurately provide the quasi-resonance modes \cite{Konoplya:2019hlu}.

\subsection{The eikonal limit}

When solving for quasinormal frequencies of black holes, various numerical schemes are applicable. Nevertheless, there is an approximation that provides a useful formula for quasinormal frequencies with great accuracy. To calculate QNMs of black holes, one can consider the so-called geometric-optics or eikonal limit as suggested by Mashhoon and Ferrari \cite{PhysRevD.31.290,PhysRevD.31.2697.2,PhysRevD.30.295}. In the eikonal limit ($\ell \to \infty$), the effective potential \eqref{veff} is simply 
\begin{align}
    V_{eik}(r) &\approx \frac{A\ell^2}{r^2}
\end{align}
This greatly simplifies the radial wave equation \eqref{radialeq}. The reduced radial equation can be solved given that the effective potential $V_{eik}$ satisfies the quantization condition 
\begin{align}
\frac{i\left(\omega^2 - V_{max}\right)}{\sqrt{-2V''_{max}}}  &= n+\frac{1}{2}.
\label{WKBeik}
\end{align}
where $V_{max}$ now denotes the maximum point of $V_{eik}$ i.e., $\frac{dV_{eik}}{dr_{\ast}}\vert_{r=r_0}=0$. It turns out that the eikonal QNMs can be expressed as the first order of the WKB formula \eqref{WKB} (e.g. see \cite{Ponglertsakul:2018smo}). The higher order terms $\Lambda_2-\Lambda_6$ can be considered as correction terms to the eikonal limit. Remarkably, it is pointed out in \cite{Cardoso:2008bp} that QNMs in the eikonal limit can be related to unstable circular null orbit around black holes in any dimensions. The real part of quasinormal frequency is determined by the angular velocity at the unstable null geodesics ($\Omega$). The imaginary part is related to the Lyapunov exponent ($\lambda_L$) which corresponds to an inverse of instabiltity timescale of the null orbit. In \cite{Hashimoto:2016dfz}, an upper bound of the Lyapunov exponent of a particle near the horizon is considered. The upper bound is determined by the surface gravity at the horizon \cite{Hashimoto:2016dfz}. Very recently, the violation of the Lyapunov exponent bound is found for Kerr-Newmann de Sitter black hole \cite{Park:2023lfc}. 

The approximation formula of the quasiormal frquencies in the eikonal limit can be expressed as \cite{Cardoso:2008bp}
\begin{align}
    \omega_{eik} &= \Omega \ell - i \left(n + \frac{1}{2}\right)|\lambda_L|, \label{weik}
\end{align}
where
\begin{align}
    \Omega &= \left. \sqrt{ \frac{A}{r^2}} \right \vert_{r=r_0}, \\
    \lambda_L &= \frac{1}{\sqrt{2}}\left.\sqrt{ A\left(A'' - \frac{6A'}{r} + \frac{6A}{r^2}\right) + A'^2  } \right \vert_{r=r_0}.
\end{align}
Both the angular velocity and the Lyapunov exponent are evaluated at $r_0$ and the prime refers to derivative with respect to $r$. We have checked and confirmed that \eqref{weik} agrees with the Pad\'e averaged WKB  in the limit $\ell \gg 1$. 

\begin{figure}[H]
    \centering
    \includegraphics[width = 8cm]{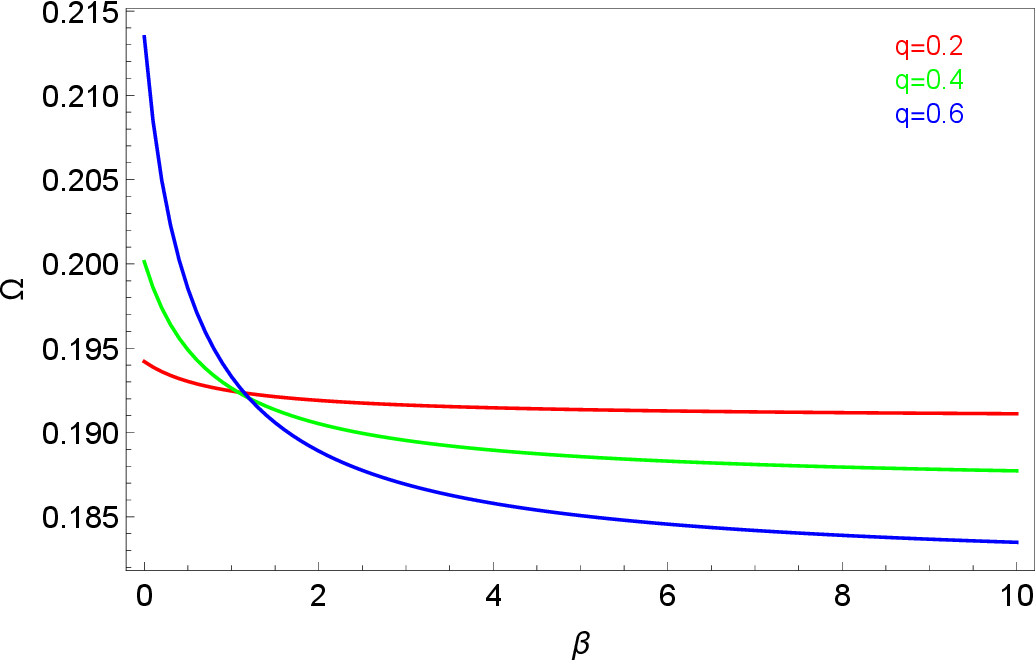}
    \includegraphics[width = 8.2cm]{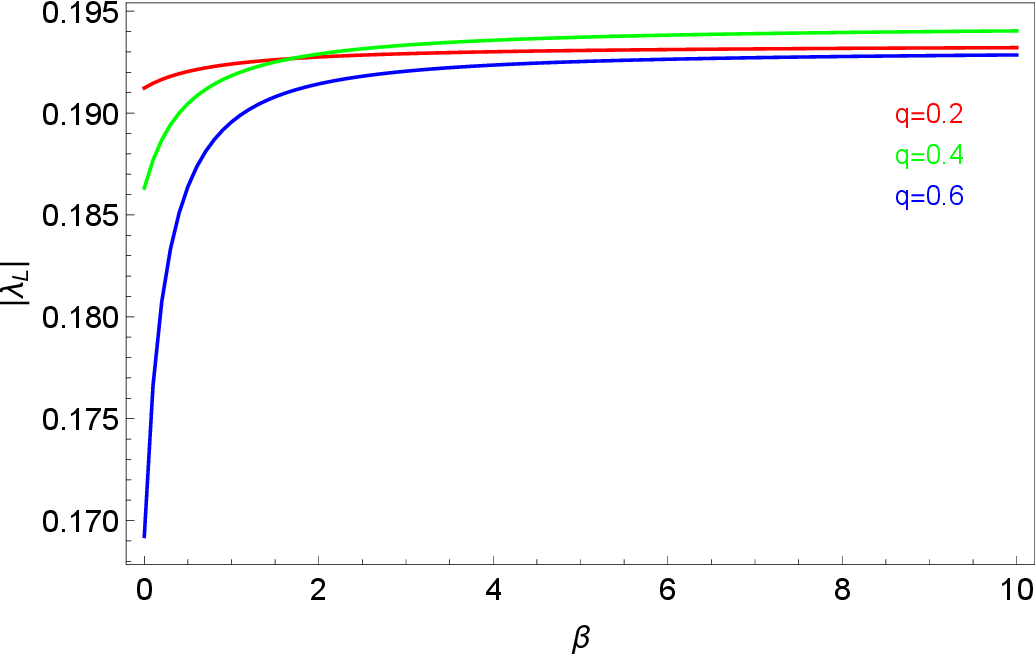}\\
    \includegraphics[width = 8cm]{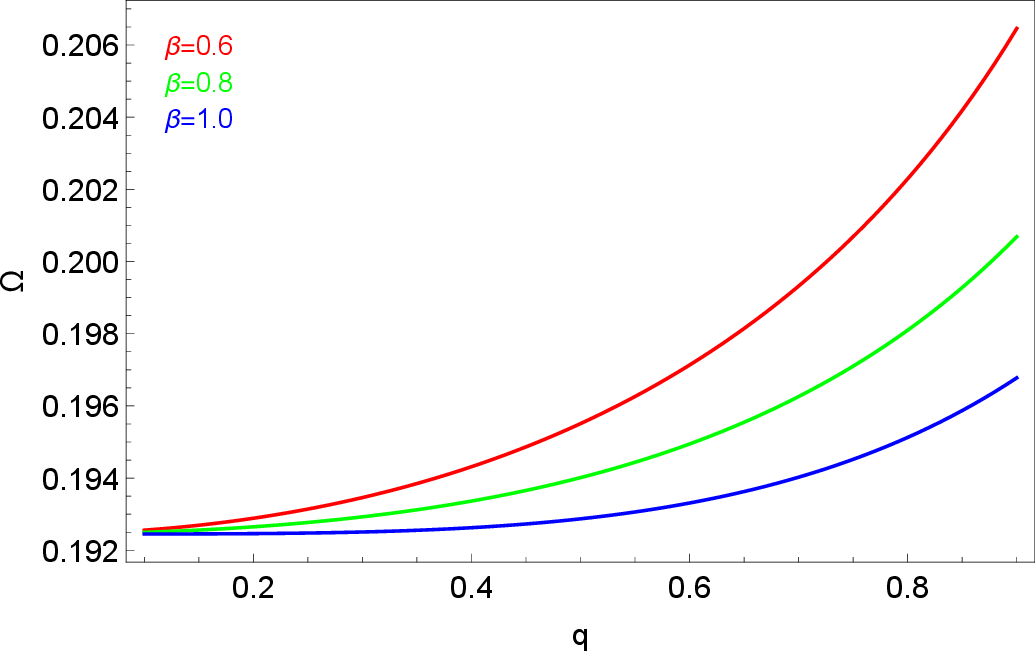}
    \includegraphics[width = 8.2cm]{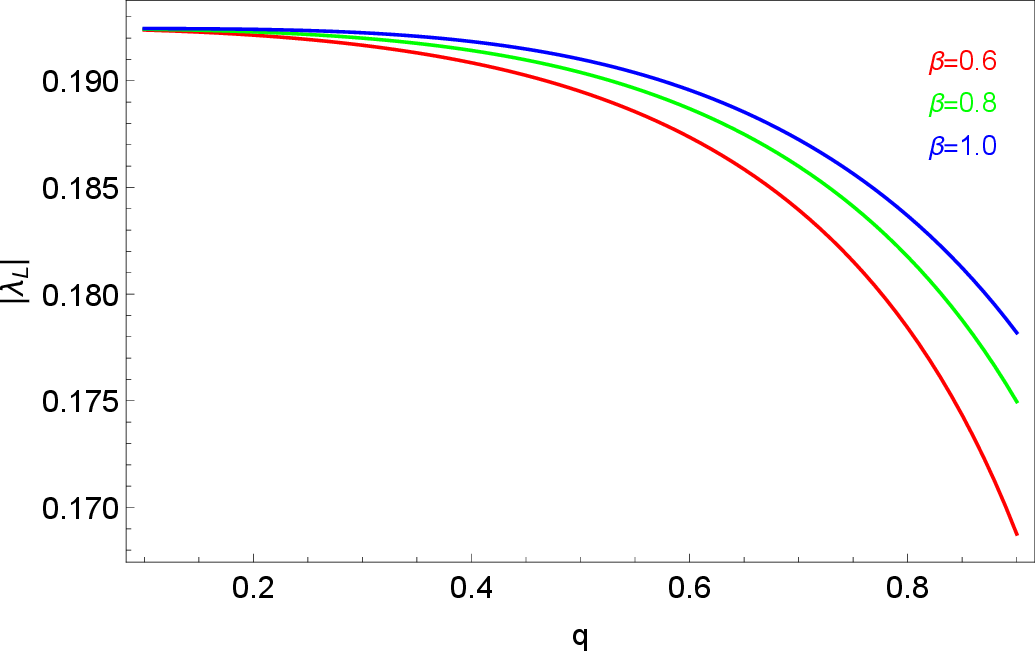}
    \caption{The angular velocity and the Lyapunov exponent as a function of $\beta$ (top) and $q$ (bottom). The black hole's mass is fixed to one. }
  \label{fig:qnm4}
\end{figure}

In Fig~\ref{fig:qnm4}, we illustrate the behaviour of the angular velocity $(\Omega)$ and the Lyapunov exponent $(\lambda_L)$ as a function of $\beta$ and $q$. As can be seen from the plots, the angular velocity decreases with $\beta$. At the lower $\beta$, the angular velocity drops significantly comparing to the larger $\beta$. Notice that, $\Omega$ changes rapidly with $\beta$ at higher black hole's charge $q$. In addition, as $\beta$ increases, the Lyapunov exponent becomes larger before approaching a certain asymptotic value. From the bottom figure, we observe that the angular velocity (the Lyapunov exponent) increases (decreases) monotonically with $q$. The behaviour of $\Omega$ against $q$ is in agreement with those found earlier in \cite{Fernando:2012yw,Panotopoulos:2019qjk} for the Reissner-Nordstrom, the Bardeen black holes and the regular black holes with exponential mass function. In contrast, our results on the Lyapunov exponent plotted against $q$ resemble the Bardeen black holes but substantially different from the Reissner-Nordstr\"om and the regular black holes \cite{Fernando:2012yw,Panotopoulos:2019qjk}. Remark that, the Lyapunov exponent of the Reissner-Nordstr\"om black hole increases with $q$ until it reaches its maximum value at some certain $q$ and then it decreases.

\section{Conclusions}\label{sec:conclud}
In this work, we study the $f(R,T)$ gravity coupled to nonlinear electrodynamics Lagrangian. With purely magnetic component of the gauge field, asymptotically flat static spherically symmetric black holes with regular centre are constructed. The black hole solutions are obtained via two approaches: I.) Fix the mass function and solve for $L_{NED}$  II.) Fix the $L_{NED}$ and solve for the mass functions. The first approach yields a functional form of a novel $L_{NED}$. Figure~\ref{fig:LNEDmassI} clearly shows the difference from the standard $U(1)$ electromagnetic Lagrangian. From the second approach, we find a generalized metric function that can be reduced to the Bardeen and the Hayward black holes in an appropriate limit. From both approaches, we find that these charged black holes possess two event horizons without essential singularities. These are shown in Fig~\ref{fig:scalarcurvature} and Fig~\ref{fig:fixmass} where the Ricci scalar and the Kretchmann scalar are plotted. 

The energy conditions (null, weak and strong) of these solutions are also explored. For the regular black holes obtained via the first approach, all the energy conditions considered here hold in exterior region of the black holes. The black hole solutions from the second approach respect the null and strong energy conditions outside the black hole's outer horizon for small value of $\beta$. As $\beta$ increases, the null and strong energy conditions are not guaranteed to hold. In contrast, the weak energy condition is violated at some certain radius inside the outer horizon for small $\beta$. As $\beta$ increases, the weak energy condition will be satisfied outside the black hole's outer horizon.

We investigate a massive scalar perturbation on these regular black holes. The corresponding quasinormal frequencies are computed via the Pad\'e average WKB method. For all cases considered in this work, the imaginary parts of the frequencies are all negative. We find that the real parts of the frequencies increase with $q,\ell$ and $\mu$ while they decrease with $\beta$. In addition, the imaginary parts of the frequencies become less negative as $q$ and $\mu$ increase, and become more negative as $\beta$ and $\ell$ increase. In the eikonal limit, the angular velocity (the Lyapunov exponent) decreases (increases) with $\beta$. Furthermore, the dependence of $\lambda_L$ on $q$ is different from those of the Reissner-Nordstr\"om black holes and the regular black holes with exponential mass function. 

There are several ways to extend this work further. Despite what the no-go theorem states in \cite{Bronnikov:2000vy}, it is crucial to show whether the $f(R,T)$ admits electrically charged regular black holes. It is interesting to consider the thermodynamics properties of regular black holes in $f(R,T)$ gravity. This problem could be challenging since the first law of thermodynamics in $f(R,T)$ is violated \cite{Jamil:2012pf}. Moreover, the photon motions around the black holes discussed in this article are also important since this could lead to the study of optical appearances of these black holes. Additionally, there are various forms of the function $f(R,T)$, and it is of great interest to explore whether they admit an exact black hole solution.

\begin{acknowledgments}
This work (Grant No. RGNS 64-217) was supported by Office of the Permanent Secretary, Ministry of Higher Education, Science, Research and Innovation  (OPS MHESI), Thailand Science Research and Innovation (TSRI) and Silpakorn university. M. Youk was supported by the Faculty of Science, Silpakorn University, Thailand, through grant SCSU-STA-2566-11.
\end{acknowledgments}


\bibliography{frt}

\end{document}